\documentclass[12pt,onecolumn,doublespace,draftclsnofoot]{IEEEtran}\IEEEoverridecommandlockouts

\usepackage{cite,graphicx,times,epsfig,amssymb,rawfonts,subfigure,array,url,setspace}
\usepackage[cmex10]{amsmath}
\newtheorem{theorem}{Theorem}[section]
\begin{document}

\title{{Cooperative Transmission in a Wireless Relay Network based on Flow
    Management} \thanks{This research is supported in part by the
    National Science Foundation under Grant CNS-0626863 and by the Air
    Force Office of Scientific Research under Grant
    FA9550-07-10456.  A part of this work was
    presented at the IEEE Wireless Communications and Networking
    Conference (WCNC), Las Vegas, March 2008.}
    \thanks{D. Chatterjee and T. F. Wong are with the Department of Electrical and Computer Engineering, University of Florida, Gainesville, FL 32611-6130, U.S.A. e-mail:\texttt{\{debdeep,twong\}@ufl.edu}}
    \thanks{T. M. Lok is with Department of Information Engineering, The Chinese University of Hong Kong, Shatin, Hong Kong e-mail:\texttt{tmlok@ie.cuhk.edu.hk}}}

\author{\emph{Debdeep Chatterjee, Tan F. Wong, and Tat M. Lok}}
\maketitle
\doublespacing

\thispagestyle{empty}

\vspace{-10mm}
\begin{abstract}
In this paper, a cooperative transmission design for a general multi-node
half-duplex wireless relay network is presented. It is assumed that the nodes
operate in half-duplex mode and that channel information is
available at the nodes. The proposed design involves solving a
convex flow optimization problem on a graph that models the relay
network. A much simpler generalized-link selection protocol based on
the above design is also presented. Both the proposed flow-optimized
protocol and the generalized-link selection protocol are shown to
achieve the optimal diversity-multiplexing tradeoff (DMT) for the relay
network. Moreover, simulation results are presented to quantify the
gap between the performances of the proposed protocols and that of a
max-flow-min-cut type bound, in terms of outage probability.
\end{abstract}

\section{Introduction}
\label{sec:intro}
A wireless relay network is one in which a set of
relay nodes assist a source node transmit information to a
destination node. Practically the wireless nodes can only support
half-duplex communication \cite{Laneman04}, i.e., no nodes can
receive and transmit information simultaneously on the same
frequency band. Different cooperative transmission schemes for
systems with half-duplex nodes have been proposed in the literature.
Fundamentally, these schemes consist of two basic steps. First, the
source transmits to the destination, and the relay listens and
``captures'' \cite{Host05} the transmission from the source at the
same time. Next, the relays send processed source information to the
destination while the source may still transmit to the destination
directly. Variants of these techniques have been proposed and have
been shown to yield good performance under different circumstances
\cite{Avestimehr06,Azarian05,Laneman04}.

Assuming channel state information (CSI) at the nodes, an
opportunistic decode-and-forward (DF) protocol for half-duplex relay
channels is proposed in \cite{Gunduz07}. 
In \cite{Ong07}, the authors present
routing algorithms to optimize the rate from a source to a
destination, based on the DF technique that uses regular block
Markov encoding and windowed decoding \cite{Xie04,Kramer03}, for the
Gaussian full-duplex multiple-relay channel. The achievable rate of
\cite{Xie04} for the Gaussian physically degraded full-duplex
multi-relay channel has been established as the capacity of this
channel in \cite{Reznik04}. In \cite{Ong06}, it is shown that the
cut-set bound on the capacity of the Gaussian single source-multiple
relay-single destination mesh network can be achieved using the
compress-and-forward (CF) method, as the relay powers go to
infinity.

Some simpler cooperative diversity methods based on network path
selection have been recently reported \cite{Bletsas06,Beres06}.
These selection methods include: (i) the max-min selection method
\cite{Bletsas06}, wherein the relay node with the maximum of the
minimum of the source-relay and relay-destination channel gains is
selected; (ii) the harmonic mean selection method \cite{Bletsas06},
wherein the relay node with the highest harmonic mean of the
source-relay and relay-destination channel gains is selected; and
(iii) the selection scheme of \cite{Beres06}, in which the relay
that can correctly decode the information from the source and has
the best relay-destination channel is selected. These methods
achieve a DMT of $d(r)=(N-1)(1-2r)$ for
an $N$ node relay network and multiplexing gain $0<r<0.5$. This is
close to what the distributed space-time coding protocol
\cite{Laneman03} achieves, when $N$ is large. Unfortunately, these
network selection protocols perform poorly in high-rate scenarios
($r>0.5$).

We have proposed a cooperative diversity design based on a flow
optimization approach for a three-node network in \cite{Wong06}. In
this design, the source node broadcasts two distinct flows to the
destination and the relay node respectively during the relay's
listen period.  Then the relay forwards this information using the
DF approach while the source may also send another flow of
information to the destination during the relay's transmit period. 
This scheme is shown to achieve the
optimal diversity order for the three-node relay channel and yield
performance very close to optimal full-duplex relaying in both
low- and high-rate situations.

Here, we apply this cooperative transmission design to a general
relay network, wherein wireless links are present between each
pair of nodes in the network.
As in \cite{Wong06}, assuming CSI is available at all nodes we use
broadcasting (BC), multiple access (MA) and time sharing (TS)
techniques to formulate a flow theoretic convex optimization problem
based on the channel conditions. Instead of considering a total
power constraint for all the transmitting nodes as in~\cite{Wong06},
we subject each node to a maximum transmit power constraint. This
yields a more reasonable system model for a general wireless
relay network, especially when the number of nodes in the relay network is
large. The resulting relaying protocol will be referred to as the
flow-optimized (FO) protocol. To obtain a more practical cooperative
design we develop a generalized-link selection (GLS) protocol, in
which we select the best relay node out of the available ones to
form an equivalent three-node relay network to transmit the information
from the source to the destination. The benefit of this, over other
network path selection strategies, becomes evident when the rate
requirement is high. It is shown that the simple GLS protocol is
optimal in terms of the DMT
\cite{Zheng03} and yields acceptable performance even when the rate
requirement is high.

Recently, in~\cite{Yuksel07}, the authors have shown that compress-and-forward (CF) relaying
achieves the optimal DMT for the three-node, half-duplex network, and that DF relaying
can achieve the optimal DMT of the four-node full-duplex network.
In this work, we show that the optimal DMT can be achieved for a general $N$-node ($N\geq3$)
half-duplex network using the FO or GLS protocols. Here, it should be clarified
that we consider that the wireless
links between each node-pair experience
independent Rayleigh fading, and this corresponds to the definition
of \emph{non-clustered} networks in~\cite{Yuksel07}.
The performances of the FO and GLS protocols are
evaluated numerically in terms of their outage probabilities for
four- and five-node relay networks for uniform and non-uniform
average power gains. The numerical results motivate the use of the
GLS protocol for situations where computation complexity is an issue
and show a remarkable improvement over the max-min selection method
of \cite{Bletsas06}. The proposed designs, based on BC and MA alone,
are sub-optimal in general. For a fair appraisal of the proposed
protocols, we compare the proposed protocols to an upper bound on
the maximum rate, derived using the max-flow-min-cut theorem
\cite[Thm.~14.10.1]{Cover91}.

\section{A General Design Using A Flow-theoretic Approach}
\label{sec:gdesgn}

We consider an $N$-node wireless relay network with a link joining each
pair of nodes. Each such wireless link is described by a bandpass
Gaussian channel with bandwidth $W$ and one-sided noise spectral
density $N_0$. We denote the power gain of the link from node $i$ to
node $j$ as $Z_{ij}$. The link power gains are assumed to be
independent and identically distributed (i.i.d.) exponential random
variables with unit mean. This corresponds to the case of
independent Rayleigh fading channels with unit average power gains.
Moreover, we assume that each node has a maximum power limit of $P$
and can only support half-duplex transmission. Note that this model
can be easily generalized to the case where channels may have
non-uniform average power gains (for which numerical examples are
presented in Section~\ref{sec:num_ex}), and where different nodes
may have different maximum power constraints. More specifically, the
latter case can be converted into the uniform maximum power
constraints case by absorbing the non-uniformity in the transmit
powers into the average power gains of the corresponding links.  In
the sequel, we characterize the system in terms of the transmit
signal-to-noise ratio (SNR), $S = \frac{P}{N_0W}$, at the input of
the links.  Time is divided into unit intervals, and BC and MA are
applied with a TS strategy that is optimized to maximize the
spectral efficiency (which we call ``rate'' hereafter for
convenience). To avoid interference between concurrent
transmissions, a time interval is divided into slots:

\begin{itemize}
\item During the first slot, the source may BC to all the other
  nodes in the network.

\item During the subsequent slots, a relay may BC to all other nodes
  (except the source node), or it may receive flows from all other
  nodes (except from the destination) through MA.

\item During the very last slot, the source and the relays may send
  information flows to the destination using MA. 
\end{itemize}
Note that the forwarding of information by the relays is based on the DF approach.
For practicality consideration, it is assumed that the phases of the
simultaneously transmitted signals from different nodes are not
synchronized. In general, for the above transmission protocol, there would be a maximum of $2(N-2)+ 2 = 2N-2$ time slots of lengths $t_1, t_2, \cdots, t_{2N-2}$ respectively.

Next, we describe the optimization problem using a graph-theoretic
formulation.  Define a graph $G=(V,E)$, where $V$ is the set of
nodes, $E$ is the set of all links joining the nodes in the graph,
and associate the vector $\underbar{r}$ to represent the flow rates
associated with each link in $E$. Thus, the number of elements in
$\underbar{r}$ equals the cardinality of $E$.  For convenience, we
write $G=(V,E,\underbar{r})$.  Now denote the source by
$\mathcal{S}$, the destination by $\mathcal{D}$, and the relay nodes
by $\mathcal{R}_1, \ldots, \mathcal{R}_{N-2}$. The slotting of a
unit time interval, as described above, yields simpler graphs for
each time slot, that we call \emph{basic graphs}. A basic graph is
either one in which a particular node may BC to several nodes, or in
which several nodes transmit via MA to a particular node. Thus for a
basic graph, we need to include only the links between the nodes
that may participate during the concerned time slot. For example,
assume that the relay $\mathcal{R}_1$ broadcasts to all nodes other than the
source, during the $i$-th time slot. The basic graph is given by
$G_i=(V,E_i,\underbar{r}_i)$ where $V=\{\mathcal{S}, \mathcal{R}_1,
\cdots, \mathcal{R}_{N-2}, \mathcal{D}\},~
E_i=\{\mathcal{R}_1\mathcal{R}_2, \cdots,
\mathcal{R}_1\mathcal{R}_{N-2}, \mathcal{R}_1\mathcal{D}\},~
\underbar{r}_i =
\left(\frac{x_{\mathcal{R}_1\mathcal{R}_2}^{i}}{t_i}~ \cdots
~\frac{x_{\mathcal{R}_1\mathcal{R}_{N-2}}^{i}}{t_i}
~\frac{x_{\mathcal{R}_1\mathcal{D}}^{i}}{t_i}\right)^T$, where
$x_{AB}^{i}$ is the flow from node $A$ to node $B$ during the $i$-th
time slot.

In general, the proposed design involves TS between the basic graphs
to yield the following equivalent graph $G$ corresponding to a unit
interval (see \cite{Wu05} for a similar idea):
\begin{equation}
G = \left( V, ~\bigcup_i E_i, ~\sum_i
t_i\underbar{r}_i\right)
= t_1 G_1 + t_2 G_2 + \ldots + t_{2N-2} G_{2N-2}.
\label{eqn:G_equiv}
\end{equation}
\noindent where the number of elements in each vector
$\underbar{r}_i$ is extended to $\vert\bigcup_i E_i\vert$ by
inserting zeros appropriately. The second equality in
(\ref{eqn:G_equiv}) implies that $G$ can be viewed as a linear
combination of the basic graphs $G_i$s, with the equivalent set of
edges given by the union of the sets $E_i$, and the equivalent flow
rate vector given by the linear combination of the individual flow
rate vectors $\underbar{r}_i$. 

To maximize the data rate from the source to the destination through
the relay network, we need to consider each cut that partitions $V$
into sets $V^s$ and $V^d$ with $\mathcal{S}\in V^s$ and
$\mathcal{D}\in V^d$.
Clearly, there can be $2^{N-2}$ such possible cuts for the $N$-node
relay network. Let these cuts and the corresponding cut sets be denoted by
$\mathcal{C}_k$, $V^s_k$, and $V^d_k$, respectively, for
$k=1,2,\cdots,2^{N-2}$.  Further, for the graph $G$, for any two
nodes $A\in V^s_k$ and $B\in V^d_k$, there exists a \emph{cut edge}
$AB$ that crosses the cut. Denote the total flow through cut edge $AB$ in a
unit time interval by $x_{AB}=\sum_{i=1}^{2N-2}x_{AB}^{i}$. Now
recall from network flow theory~\cite{Ahuja93} that the maximum flow
rate from the source to the destination is specified by the minimal
cut of the equivalent graph (\ref{eqn:G_equiv}). Consequently, we
arrive at the following convex flow optimization problem that can be
solved using standard optimization techniques:
\begin{equation}
  \max\min \left(\sum_{A\in V^s_1,B\in V^d_1}x_{AB}, \sum_{A\in
      V^s_2,B\in V^d_2}x_{AB}, \cdots, \sum_{A\in V^s_{2^{N-2}},B\in
      V^d_{2^{N-2}}}x_{AB} \right)\label{eqn:main_opt_prob1}
\end{equation}
\noindent over all flow allocations $x_{AB}^{i}$ and all time slot
lengths $t_i$, subject to
\begin{itemize}
\item the \emph{non-negativity constraints}: $x_{AB}^{i}$, $t_i \geq
  0$ for all cut edges $AB$ and $i=1,2,\cdots,2N-2$,
\item the \emph{total-time constraint}: $t_1+\ldots+t_{2N-2} =1$,
\item the \emph{power (capacity) constraints}:
\begin{itemize}
\item for a BC slot the flow rates should lie in the capacity region
  of the BC channel with the transmitting node having a power
  constraint of $P$,
\item for an MA slot the flow rates should lie in the capacity region
  of the MA channel with a maximum power constraint $P$ for each
  transmitting node,
\end{itemize}
\item the \emph{flow constraints}: considering steady state operation,
  the total information flow out of a relay should equal the flow into
  the relay in each unit time interval.
\end{itemize}

\noindent Note that the dependence of the objective function on the channel gains and the
time slot lengths is implicitly expressed through the capacity constraints.
Denote the cut separating $\mathcal{S}$ from all the other nodes and
the cut separating $\mathcal{D}$ from all nodes as
$\mathcal{C}_{\mathcal{S}}$ and $\mathcal{C}_{\mathcal{D}}$,
respectively.  Then we observe that the cost function in
(\ref{eqn:main_opt_prob1}) above can be further simplified
to $\max~\min\left\{x(\mathcal{C}_{\mathcal{S}}),~x(\mathcal{C}_{\mathcal{D}})\right\}$,
where
\begin{equation}
\displaystyle x(\mathcal{C}_{\mathcal{S}}) = x_{\mathcal{SD}} +
\sum_{j=1}^{N-2}x_{\mathcal{S}\mathcal{R}_j}\mathrm{~~~ and ~~~}
x(\mathcal{C}_{\mathcal{D}}) = x_{\mathcal{SD}} +
\sum_{i=1}^{N-2}x_{\mathcal{R}_i\mathcal{D}}\label{eqn:BC_cut}
\end{equation}
are the total flows across the above-mentioned cuts
$\mathcal{C}_{\mathcal{S}}$ and $\mathcal{C}_{\mathcal{D}}$,
respectively.  To see this, consider the cut $\mathcal{C}$ with
$V^s=\{\mathcal{S},\mathcal{R}_1,\cdots,\mathcal{R}_l\}$, and
$V^d=\{\mathcal{R}_{l+1},\cdots,\mathcal{R}_{N-2},\mathcal{D}\}$ for
some $l\in\{1,2,\cdots,N-2\}$. The total flow across this cut is
given by
\begin{equation}
\displaystyle x(\mathcal{C}) = x_{\mathcal{SD}} +
\sum_{j=l+1}^{N-2}x_{\mathcal{S}\mathcal{R}_j} +
\sum_{i=1}^l\left(x_{\mathcal{R}_i\mathcal{D}} +
\sum_{j=l+1}^{N-2}x_{\mathcal{R}_i\mathcal{R}_j}\right).\label{eqn:flow_thru_C}
\end{equation}

\noindent Now, consider node $i$ for $i\in\{1,2,\cdots,l\}$.
According to the flow constraint for node $i$,
\begin{equation}
\displaystyle x_{\mathcal{R}_i\mathcal{D}} +
\sum_{j=l+1}^{N-2}x_{\mathcal{R}_i\mathcal{R}_j} + \sum_{k=1,k\neq
i}^{l}x_{\mathcal{R}_i\mathcal{R}_k} = x_{\mathcal{S}\mathcal{R}_i}
+ \sum_{j=l+1}^{N-2}x_{\mathcal{R}_j\mathcal{R}_i} + \sum_{k=1,k\neq
i}^{l}x_{\mathcal{R}_k\mathcal{R}_i}.\label{eqn:node_i_flow}
\end{equation}
Summing (\ref{eqn:node_i_flow}) over all $i\in\{1,2,\cdots,l\}$ we get
\begin{equation}
\sum_{i=1}^l\left(x_{\mathcal{R}_i\mathcal{D}}
+ \sum_{j=l+1}^{N-2}x_{\mathcal{R}_i\mathcal{R}_j}\right)=
\sum_{i=1}^l\left(x_{\mathcal{S}\mathcal{R}_i} +
\sum_{j=l+1}^{N-2}x_{\mathcal{R}_j\mathcal{R}_i}\right).\label{eqn:nodes_i2l_flow}
\end{equation}
Since
$\sum_{i=1}^l\sum_{j=l+1}^{N-2}x_{\mathcal{R}_j\mathcal{R}_i}\geq 0$,
combining (\ref{eqn:BC_cut}), (\ref{eqn:flow_thru_C}) and
(\ref{eqn:nodes_i2l_flow}) gives $x(\mathcal{C}) \geq
x(\mathcal{C}_{\mathcal{S}})$.  Similarly, we have
$x(\mathcal{C}) \geq x(\mathcal{C}_{\mathcal{D}})$. Thus the cost
function in (\ref{eqn:main_opt_prob1}) reduces to the above-mentioned form.

\section{Generalized-link Selection and Its Optimality}\label{sec:optimality}

In this section, we present the GLS protocol and establish the
optimality of the FO and GLS protocols in terms of the
DMT.  This is accomplished in three
steps. First, we apply the FO protocol to the three-node relay network.
Next, we propose the GLS protocol based on a selection strategy that
is sub-optimal to the FO protocol of Section~\ref{sec:gdesgn}.
Finally, the optimality of the GLS protocol, and thereby, that of the FO
protocol, is established.

\subsection{The Three-node Relay Network} \label{subsec:3node}

The three-node relay network consists of a source ($\mathcal{S}$), a relay
($\mathcal{R}$), and a destination ($\mathcal{D}$). We specialize
the general design described in the previous section to this
three-node relay network.  A unit time interval is divided into two time
slots of lengths $t_1$ and $t_2$ with $t_1 + t_2 = 1$.
During the first time slot, $\mathcal{S}$ sends (via BC) two flows
of rates $x_{\mathcal{SD}}^{1}/t_1=x_1/t_1$ and
$x_{\mathcal{SR}}^{1}/t_1=x_2/t_1$ to $\mathcal{D}$ and
$\mathcal{R}$, respectively, resulting in the basic graph $G_1$. 
During the second time slot, $\mathcal{R}$ and $\mathcal{S}$ send
(via MA) two flows of rates $x_{\mathcal{RD}}^{2}/t_2=x_4/t_2$ and
$x_{\mathcal{SD}}^{2}/t_2=x_3/t_2$ to $\mathcal{D}$, respectively,
resulting in the basic graph $G_2$. 
Combining the two basic graphs yields the
equivalent graph as $G = t_1G_1 + t_2G_2$. Note that the information
flow of rate $x_4/t_2$ sent by $\mathcal{R}$ during the MA time slot
is from the flow of rate $x_2/t_1$ it received during the BC time
slot. 
Thus, we have the flow constraint $x_4=x_2$. The rate for this
network is specified by the min-cut which is clearly $\min\{(x_1 +
x_2 + x_3),(x_1+x_4+x_3)\}$. Hence, the flow optimization problem is
given by:
\begin{equation}
\max~\min~\{(x_1 + x_2 + x_3),(x_1+x_4+x_3)\} \label{eqn:3node_opt}
\end{equation}
over flow allocations $x_1,~x_2,~x_3,~x_4$, and time slot lengths
$t_1,~t_2$, subject to
\begin{itemize}
\item {\it non-negativity constraints: } $x_1, x_2, x_3, x_4 \geq
  0,\quad t_1, t_2 \geq 0$,
\item {\it total-time constraint: } $t_1+t_2=1$,
\item {\it power constraints: } $S_{BC}\leq S,~x_1\leq t_1C(Z_{\mathcal{SD}}S),~x_2\leq t_1C(Z_{\mathcal{SR}}S)$ for the BC slot,\\
  $\qquad x_3\leq t_2C(Z_{\mathcal{SD}}S),~ x_4\leq
  t_2C(Z_{\mathcal{RD}}S),~ x_3 + x_4\leq
  t_2C(Z_{\mathcal{SD}}S+Z_{\mathcal{RD}}S)$ for the MA slot,
\item {\it flow constraint: } $x_2 = x_4$,
\end{itemize}
where $C(x)=\log(1+x)$,  and $S_{BC}$, the minimum SNR required for
the source to broadcast at rates $x_1/t_1$ and $x_2/t_1$ to the
destination and the relay, respectively, in the first time slot with
$0<t_1\leq1$, is given by (see~\cite[Lemma 3.1]{Wong06} for proof)
\[
S_{BC}= \left \{
\begin{array}{ll}
\frac{1}{Z_{\mathcal{S}\mathcal{D}}} ( e^{x_1/t_1} -1) +
\frac{1}{Z_{\mathcal{S}\mathcal{R}}} e^{x_1/t_1} ( e^ {x_2/t_1} -1)
\mathrm{~ for ~} Z_{\mathcal{S}\mathcal{R}}> Z_{\mathcal{S}\mathcal{D}}, \\
\frac{1}{Z_{\mathcal{S}\mathcal{R}}} ( e^{x_2/t_1} -1) +
\frac{1}{Z_{\mathcal{S}\mathcal{D}}} e^{x_2/t_1} ( e^ {x_1/t_1} -1)
\mathrm{~ for ~} Z_{\mathcal{S}\mathcal{R}} \leq
Z_{\mathcal{S}\mathcal{D}}.
\end{array}
\right .
\]
For $t_1=0$, $S_{BC}=0$. Note that for the BC slot, the last two constraints are redundant when $t_1>0$, and complements the first constraint when $t_1=0$.

The solution of this flow optimization problem is given in
Appendix~\ref{app:opt_soln}. As mentioned in Section~\ref{sec:intro},
the above optimization problem formulation is different from that
in~\cite{Wong06} wherein the sum of the source and relay powers,
required to achieve a certain data rate, is minimized. More
specifically, when considering individual power constraints for each
node, we cannot use part $2$ of~\cite[Lemma 3.1]{Wong06} to describe
the power constraints for the MA slot. This is because doing so would
restrict the flows $x_2$ and $x_3$ such that the sum of powers
expended at $\mathcal{S}$ and $\mathcal{R}$ is minimized.  On the
other hand, in the present problem, the power constraints only dictate
that the flow-rates should lie in the MA capacity region specified by
the maximum power available at each transmitting node, for the
particular fading state. With this modification in the constraint for
the MA slot, the solution approach to the above problem needs to be
markedly different from that in~\cite{Wong06} as shown in Appendix~\ref{app:opt_soln}.
The maximum information rate from the source $\mathcal{S}$ to the
destination $\mathcal{D}$ for different cases is summarized below:

\noindent a) $Z_{\mathcal{SD}}\geq Z_{\mathcal{SR}}$: The maximum
rate is $X(S) = C(Z_{\mathcal{SD}}S)$ with direct transmission
from $\mathcal{S}$ to $\mathcal{D}$. 

\noindent b) $Z_{\mathcal{SD}}<Z_{\mathcal{SR}}$: The maximum rate
is $X(S) = \max_{0\leq t_2\leq t_{2\textnormal{max}}}
t_1\log\left(\frac{1+Z_{\mathcal{SD}}S}
{1+\frac{Z_{\mathcal{SD}}}{Z_{\mathcal{SR}}}\left[\left(1+\frac{Z_{\mathcal{RD}}S}{1+Z_{\mathcal{SD}}S}\right)^{t_2/t_1}-1\right]}\right)
+ t_2C(Z_{\mathcal{SD}}S+Z_{\mathcal{RD}}S)$ with
$t_{2\textnormal{max}} =
C(Z_{\mathcal{SR}}S)/\left[C(Z_{\mathcal{SR}}S)+C(Z_{\mathcal{RD}}S+Z_{\mathcal{SD}}S)-C(Z_{\mathcal{SD}}S)\right]$
and $t_1 = 1 - t_2$.

Thus for a given power limit (i.e. a given $S$) at the nodes,
relaying is advantageous only when
$Z_{\mathcal{SD}}<Z_{\mathcal{SR}}$. Further, the optimal solution
always allocates a non-zero flow to the direct link. Also, the
relay-destination link gain $Z_{\mathcal{RD}}$ does not influence
the strategy of transmission (i.e. whether to use only the direct
link or both the relay and direct links), but only the amount of
information through the relay link.

\subsection{Generalized-link Selection} \label{subsec:sub_opt}
For the general $N$-node relay network, the flow optimization solution can
be computationally demanding even for moderate values of $N$.
The GLS protocol described below provides a simple sub-optimal
design to address this complexity issue.  In essence, the GLS
protocol identifies the \emph{best} relay path out of the possible
$N-2$ relay paths and considers only the chosen relay along with the
source and destination to form a three-node relay network, which we call a
\emph{generalized-link} from the source to the destination, for
information transmission. In other words, the aim is to choose the
best relay such that the equivalent three-node relay network obtained
(containing the source, destination and the chosen relay) gives the
maximum rate over all possible equivalent three-node networks
containing the source and destination.  More precisely, we need to
consider the following possibilities:
\begin{enumerate}
\item $Z_{\mathcal{SD}}\geq Z_{\mathcal{S}\mathcal{R}_i}$ for all
  $i\in I = \{1,2,\cdots,N-2\}$: From the results of the optimization
  problem (\ref{eqn:3node_opt}), it is clear that the maximum rate
  would be $C(Z_{\mathcal{SD}}S)$ with direct transmission of all data
  from the source to the destination without using any relay.

\item There exists a $k\in I$ such that
  $Z_{\mathcal{S}\mathcal{R}_k}>Z_{\mathcal{SD}}$: Let the set of all
  such node indices be $K$ and for all $i\in I\setminus K$,
  $Z_{\mathcal{SD}}\geq Z_{\mathcal{S}\mathcal{R}_i}$. For this case,
  choose the node $\mathcal{R}_k'$ as the relay such that
  $k'=\arg\max_{k\in K} X_k(S)$, where $X_k(S)$ is the maximum rate
  for the three-node relay network with the source $\mathcal{S}$, the relay
  $\mathcal{R}_k$ and destination $\mathcal{D}$.
\end{enumerate}
In terms of the worst-case computational complexities for the FO and GLS
protocols, it can be seen that, for an $N$-node relay network with $N>3$, the FO protocol
involves a max-min optimization over $2(N^2-2N+2)$ variables (all possible flows and time slot
lengths), subject to $N-1$ non-linear and $2(N^2-N+1)$ linear constraints, whereas
the GLS protocol involves a maximum of $N-2$ maximizations of a non-linear concave function over two variables,
subject to two linear constraints, followed by finding the maximum of $N-2$ real numbers with
a worst-case complexity of $O(N-2)$. Moreover, for $N>3$, for the FO protocol, the BC slots
potentially involve $(N-1)$- and $(N-2)$-level superposition coding (SPC) or dirty paper coding (DPC)
implementations for $\mathcal{S}$ and the relays respectively, while the MA slots at
the relays and $\mathcal{D}$ may involve a maximum of $(N-3)$ and $(N-2)$
interference cancelation (IC) operations respectively. On the other hand, the GLS protocol
involves a maximum of $2$-level SPC/DPC and one IC operation for the BC and MA slots respectively,
for any $N>3$.

\subsection{Diversity-multiplexing tradeoff} \label{subsec:dm_tradeoff}

As in \cite{Zheng03}, the multiplexing gain $r = \lim_{S \rightarrow
  \infty} \frac{R(S)}{\log S}$ where $S$ is the SNR and $R(S)$ is the
rate at an SNR level of $S$.
Following \cite{Zheng03}, we parameterize the system, in terms of the
SNR $S$ and the multiplexing gain, $0<r<1$, with the rate increasing
with the SNR as $R = r\log(S)$.  With the parameterization $(r,S)$,
the diversity order achieved by the transmission scheme is given by
\begin{equation}
d(r) = \lim_{S \rightarrow \infty} \frac{-\log P_{e}
(r,S)}{\log S}\label{eqn:div_def}
\end{equation}
where $P_{e}(r,S)$ is the average probability of error when the SNR is
$S$ and multiplexing gain $r$.
The following theorem, whose proof is outlined in
Appendix~\ref{app:pf_dm_sub_opt}, establishes the optimality of the
the GLS protocol (and hence the FO protocol) in terms of the
DMT:
\begin{theorem} \label{thm:dm_sub_opt}
  The GLS and FO protocols achieve the optimal DMT
  $d(r)=(N-1)(1-r)$ for all $0<r<1$, for the $N$-node
  half-duplex wireless relay network.
\end{theorem}

\section{Numerical Examples}\label{sec:num_ex}

Using the outage probability as the performance metric, we compare
the FO and GLS protocols against the max-min selection method of
\cite{Bletsas06}, as it provides the best performance amongst
previously proposed path selection methods, and an outage
probability lower bound derived using the max-flow-min-cut theorem
of \cite[Thm. 14.10.1]{Cover91}.

\begin{figure}[t]
\begin{center}
\includegraphics[scale=0.8]{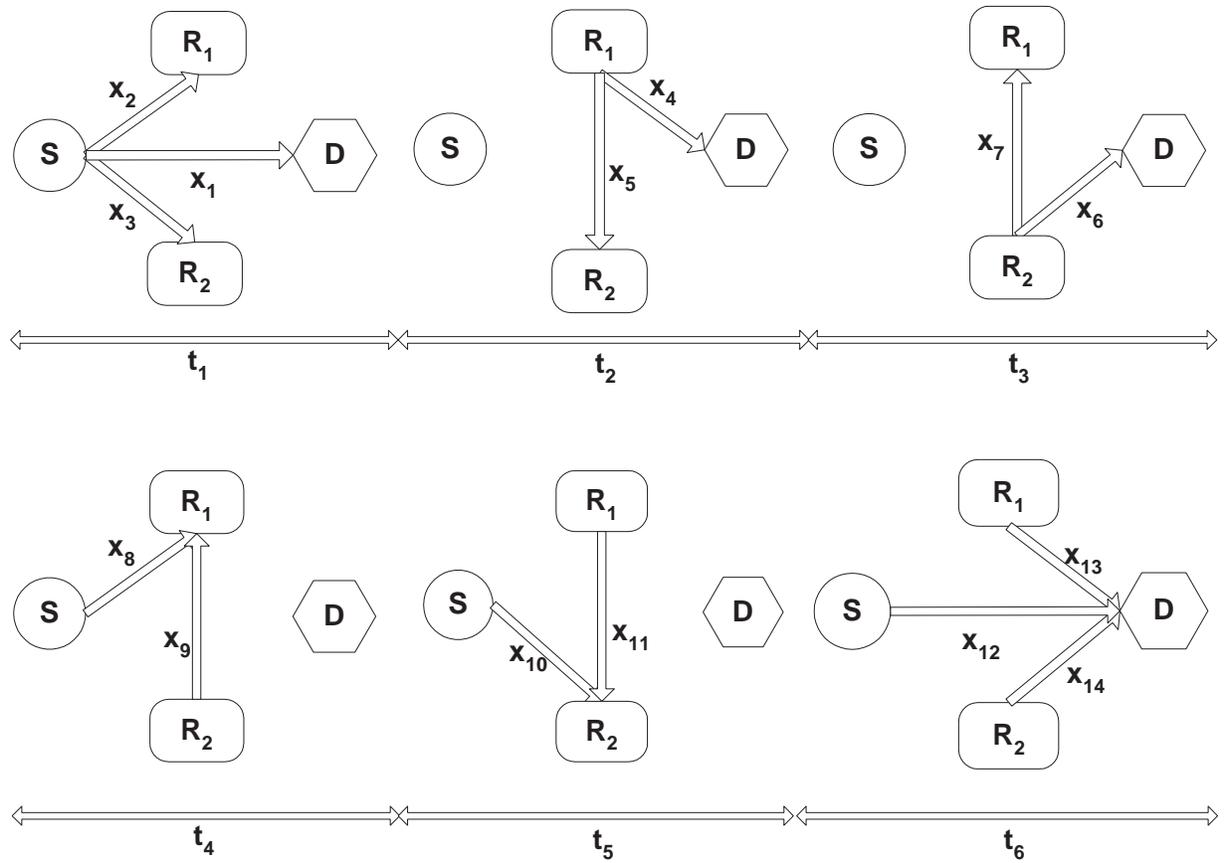}
\end{center}
\caption{FO protocol for the four-node relay network with
$t_1+\cdots+t_6=1$. The flow optimization is performed over all
flows $x_1,~\cdots,~x_{14}$, and all time slot lengths
$t_1,~\cdots,~t_6$.} \label{fig:4_node_FO}
\end{figure}

\begin{figure}[b]
\begin{center}
\includegraphics[scale=0.8]{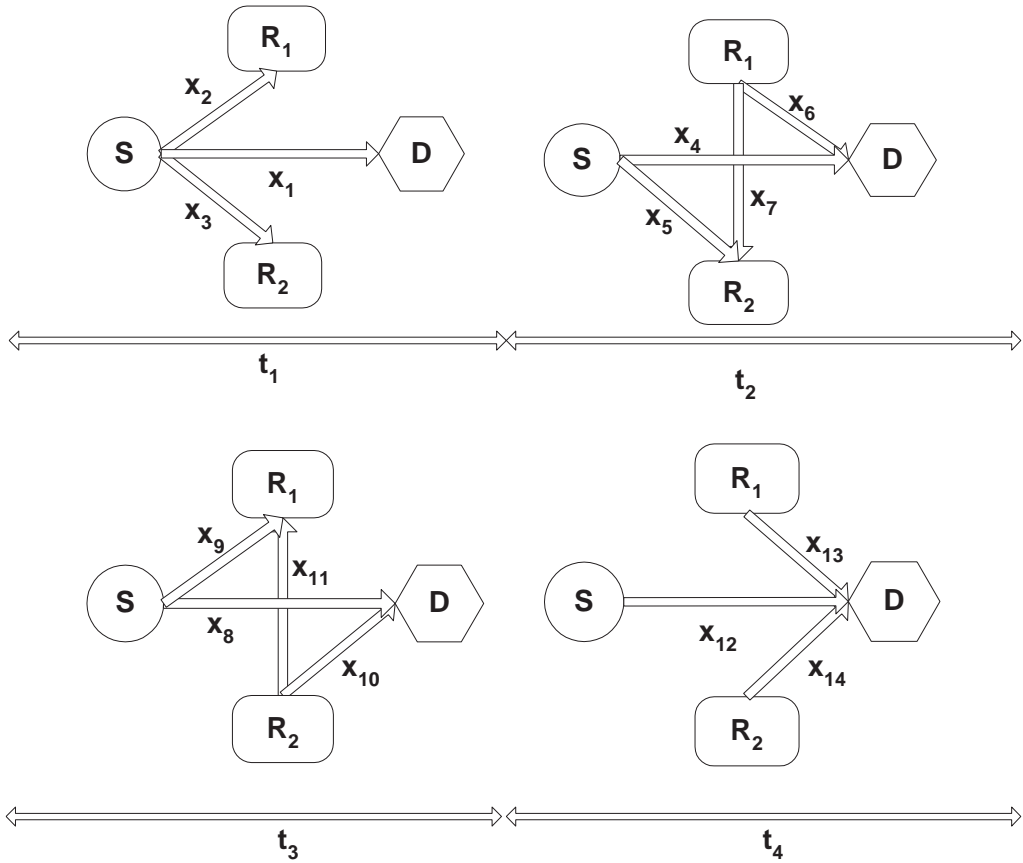}
\end{center}
\caption{Transmission strategy to obtain a lower bound on the outage
probability for the four-node relay network. Here $t_1+\cdots+t_4=1$, and
the optimization is over $x_1,~\cdots,~x_{14}$, and
$t_1,~\cdots,~t_4$, with the application of the max-flow-min-cut
theorem for the intermediate slots.} \label{fig:4_node_lb}
\end{figure}

For the four-node relay network, there can be 6 possible time slots in the
FO protocol as shown in Fig.~\ref{fig:4_node_FO}.
To derive an upper bound on the achievable rate (and thereby a lower
bound on the outage probability), we use max-flow-min-cut type
bounds for half-duplex communication. There are four possible time
slots as shown in Fig.~\ref{fig:4_node_lb}, with the first BC slot
and the last MA slot at the destination same as in the FO protocol,
but now, the source and a relay may transmit simultaneously to the
other relay and the destination during each of the intermediate
slots over interference channels. We use the max-flow-min-cut
theorem to upper bound the maximum information flow in these two
time slots.

For the five-node relay network, there can be 8 possible time slots in the
FO protocol - four BC slots for the source and the three relays to
transmit information, and four MA slots for the three relays and the
destination to receive information respectively. Similar to the
four-node relay network case, for the max-flow-min-cut bound, there are
$8$ possible time slots with the first BC slot and the last MA slot
at the destination being the same as for the FO protocol, and
multi-source-multi-destination transmissions during the six
intermediate slots.

\begin{figure}[t]
\begin{center}
\includegraphics[scale=1.0]{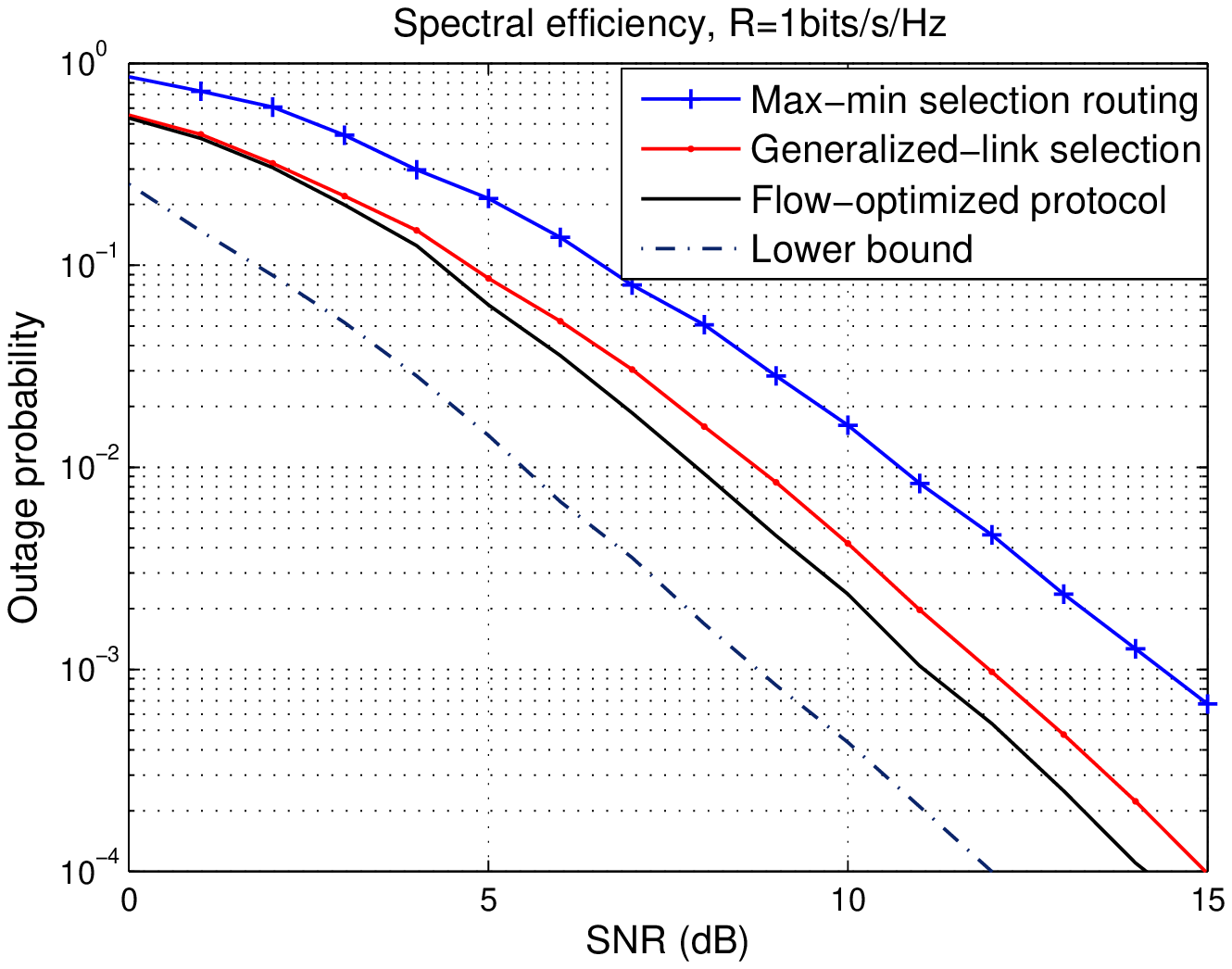}
\end{center}
\caption{Four-node relay network with uniform average power gains:
Outage probabilities for required rate $R=1$bit/s/Hz.}
\label{fig:case_1u_R1}
\end{figure}
\begin{figure}[t]
\begin{center}
\includegraphics[scale=1.0]{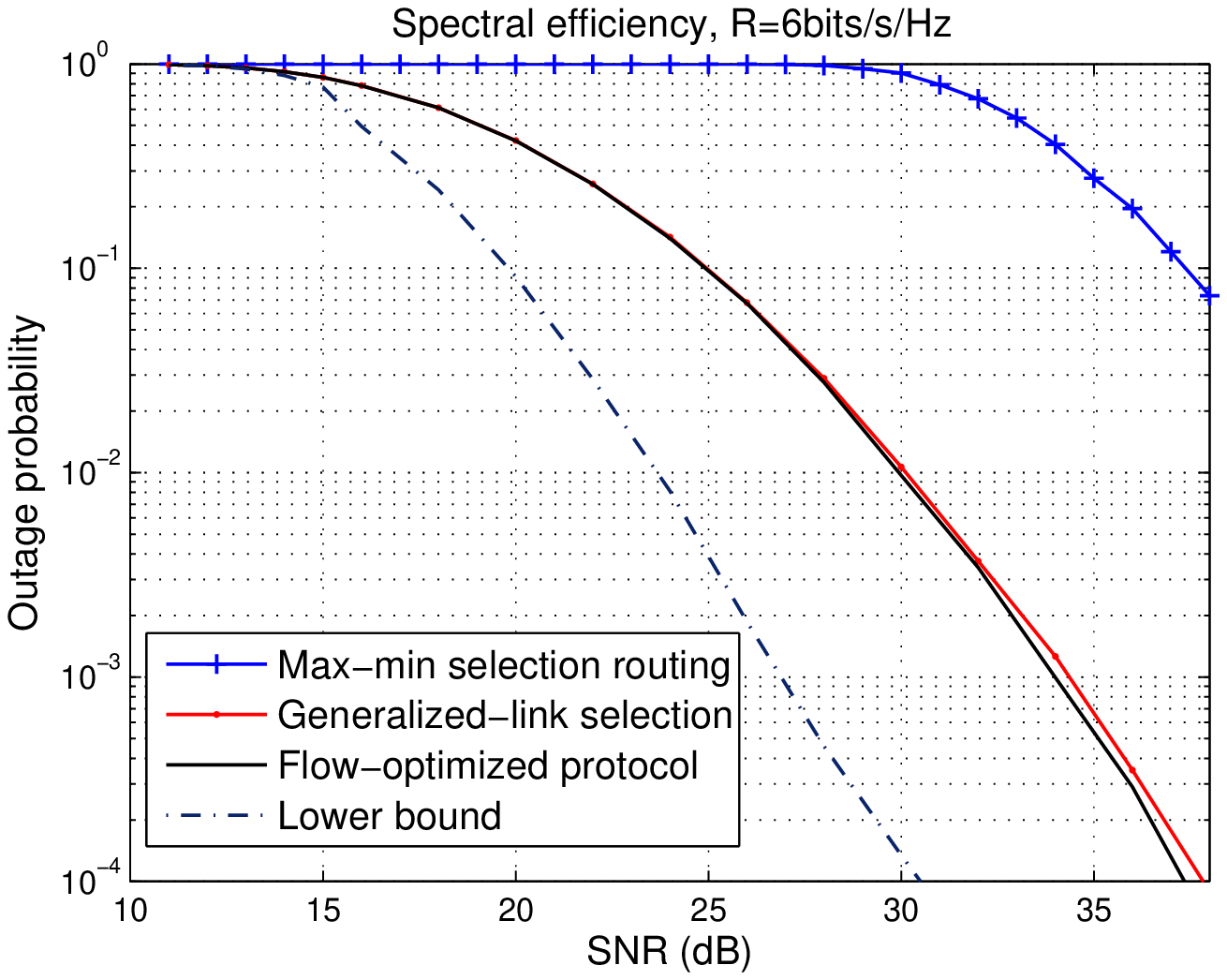}
\end{center}
\caption{Four-node relay network with uniform average power gains:
Outage probabilities for required rate $R=6$bits/s/Hz.}
\label{fig:case_6u_R6}
\end{figure}
\begin{figure}[t]
\begin{center}
\includegraphics[scale=1.0]{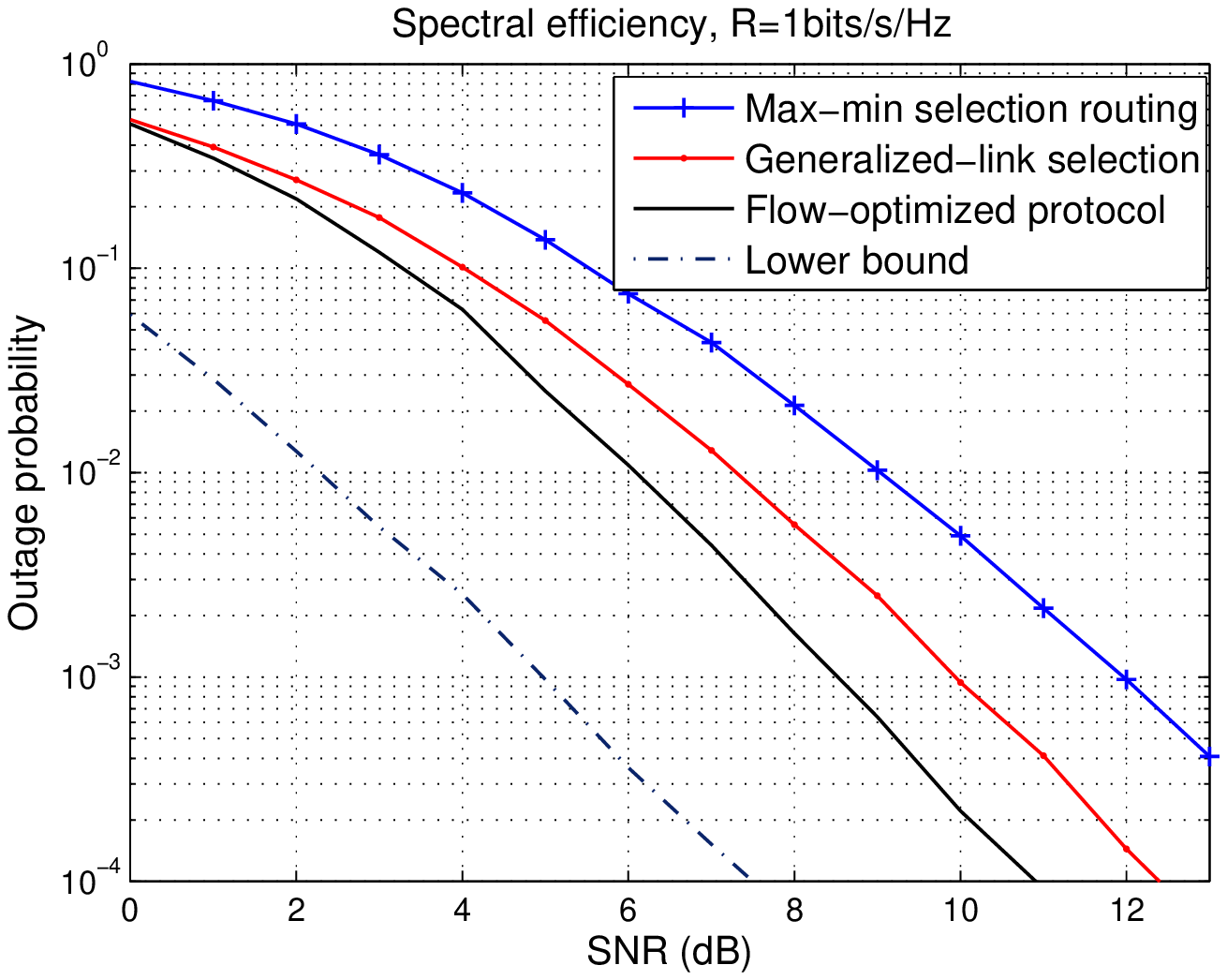}
\end{center}
\caption{Five-node relay network with uniform average power gains:
Outage probabilities for required rate $R=1$bits/s/Hz.}
\label{fig:5node_case_1u_R1}
\end{figure}
\begin{figure}[t]
\begin{center}
\includegraphics[scale=1.0]{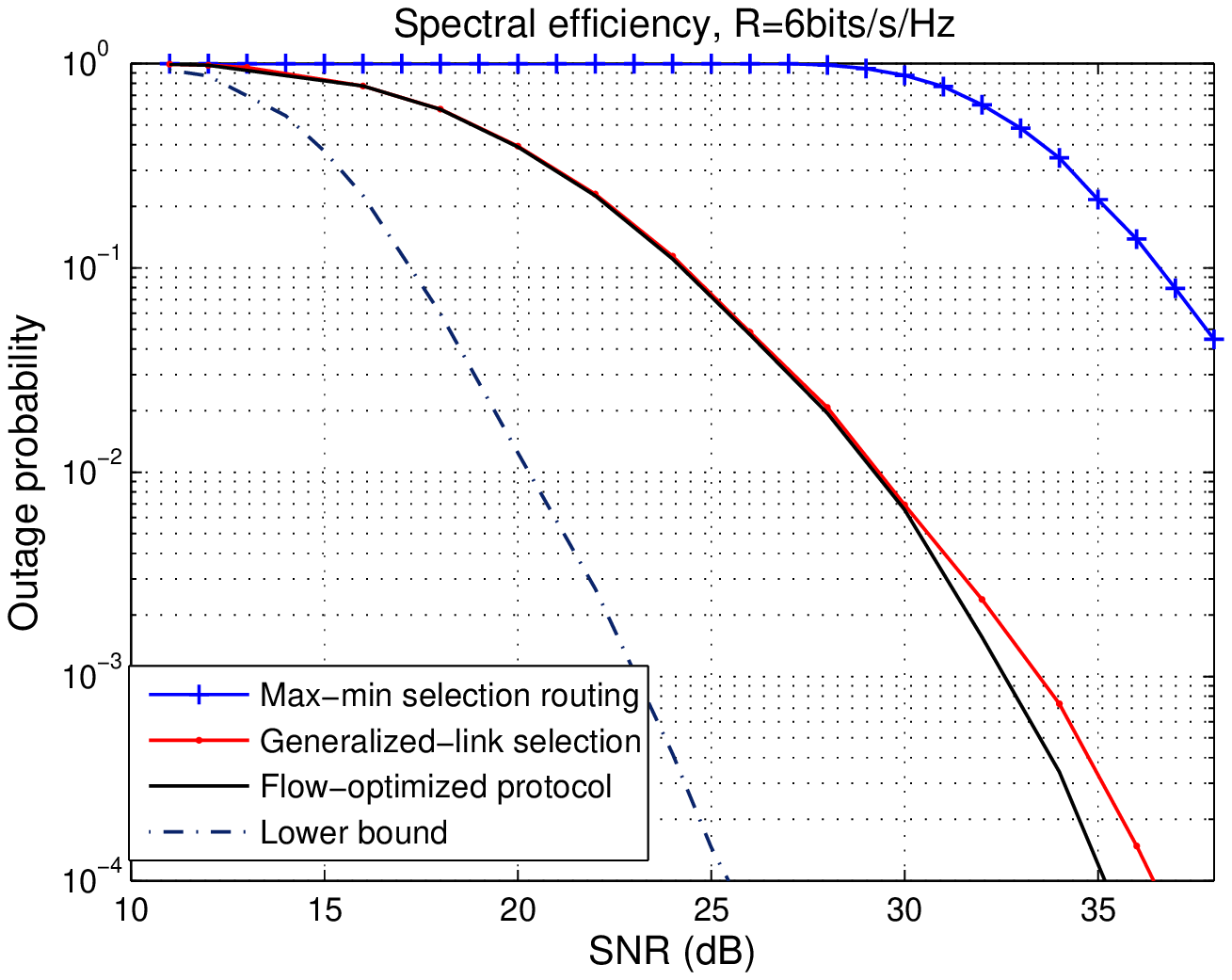}
\end{center}
\caption{Five-node relay network with uniform average power gains:
Outage probabilities for required rate $R=6$bits/s/Hz.}
\label{fig:5node_case_6u_R6}
\end{figure}

With the above division of time slots, the formalization of the
problem is done as in the previous sections, and we use the
optimization routine of \cite{Lawrence97} to obtain the maximum
achievable rates and upper bounds for different values of required
rates. In Figs.~\ref{fig:case_1u_R1} and \ref{fig:case_6u_R6}, we
plot the outage probabilities of the various schemes with the
required rate $R$ at $1$bit/s/Hz and $6$bits/s/Hz respectively, for
the four-node relay network. Figs.~\ref{fig:5node_case_1u_R1} and
\ref{fig:5node_case_6u_R6} present the same for the five-node
relay network. When compared to the FO protocol, the GLS protocol suffers
a loss of around  $1.0$dB, and around $1.5$dB (when $R$ is either
$1$bit/s/Hz or $6$bits/s/Hz), at an outage probability of $10^{-4}$,
for the four- and five-node relay networks respectively. On the other
hand, the performance degradation for the max-min selection method
of \cite{Bletsas06}, as compared to the FO protocol or even the GLS
protocol, is more than $12$dB at an outage probability of
$7.0\times10^{-2}$, when $R=6$bits/s/Hz for the four-node relay network,
and an exactly similar situation can be observed for the five-node
relay network. Moreover, for the four-node relay network, the FO protocol is
within $2.14$dB (when $R=1$bit/s/Hz) to within $7.05$dB (when
$R=6$bits/s/Hz) of the lower bound on the outage probability when
the outage probability is $10^{-4}$. For the five-node relay network, the
corresponding differences are approximately $3$dB and $9.6$dB
respectively. 

\begin{figure}[t]
\begin{center}
\includegraphics[scale=1.0]{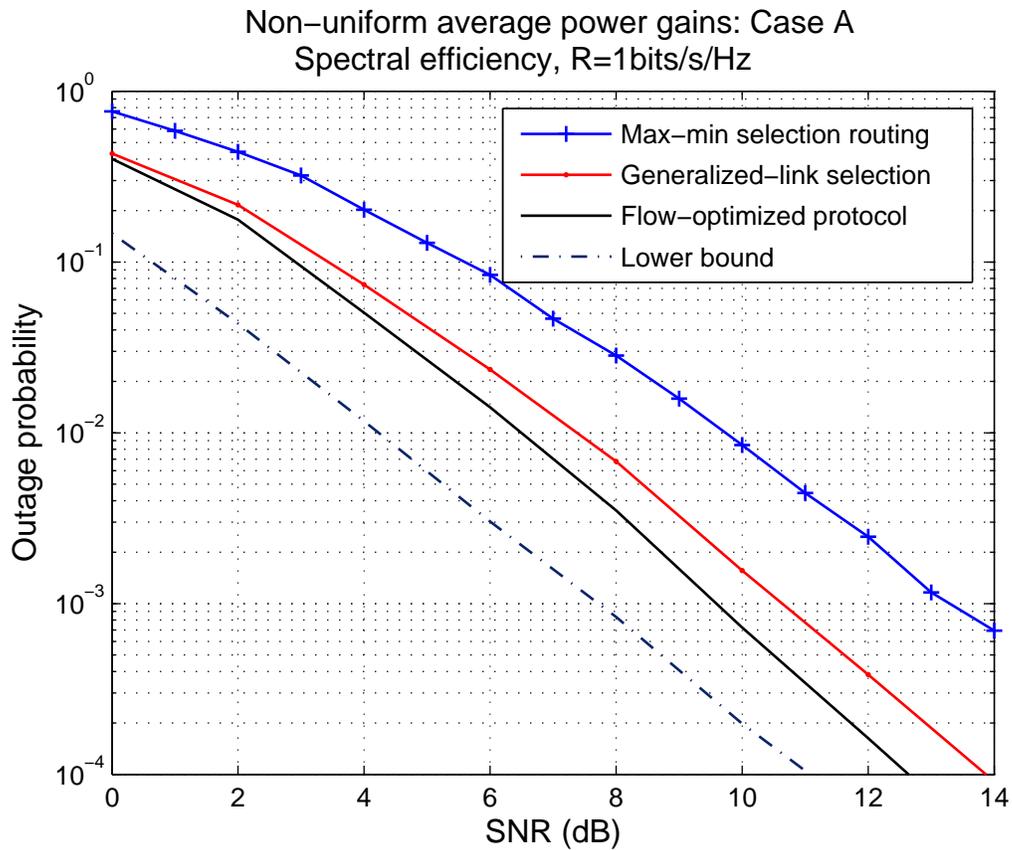}
\end{center}
\caption{Four-node relay network with non-uniform average power
gains.  Case A:
$\mathbb{E}[Z_{\mathcal{SR}_1}]=2.0,~\mathbb{E}[Z_{\mathcal{SR}_2}]=2.0,~\mathbb{E}[Z_{\mathcal{SD}}]=1.0,
~\mathbb{E}[Z_{\mathcal{R}_1\mathcal{R}_2}]=1.0,~\mathbb{E}[Z_{\mathcal{R}_1\mathcal{D}}]=1.5,~\mathbb{E}[Z_{\mathcal{R}_2\mathcal{D}}]=1.0$.
In this case, both the source-relay links
are, on average, better than the direct link, and one
relay-destination link is, on average, better than the other,
resulting in relay $\mathcal{R}_1$ being more favorable than relay $\mathcal{R}_2$.}
\label{fig:case_1a_R1}
\end{figure}
\begin{figure}[t]
\begin{center}
\includegraphics[scale=1.0]{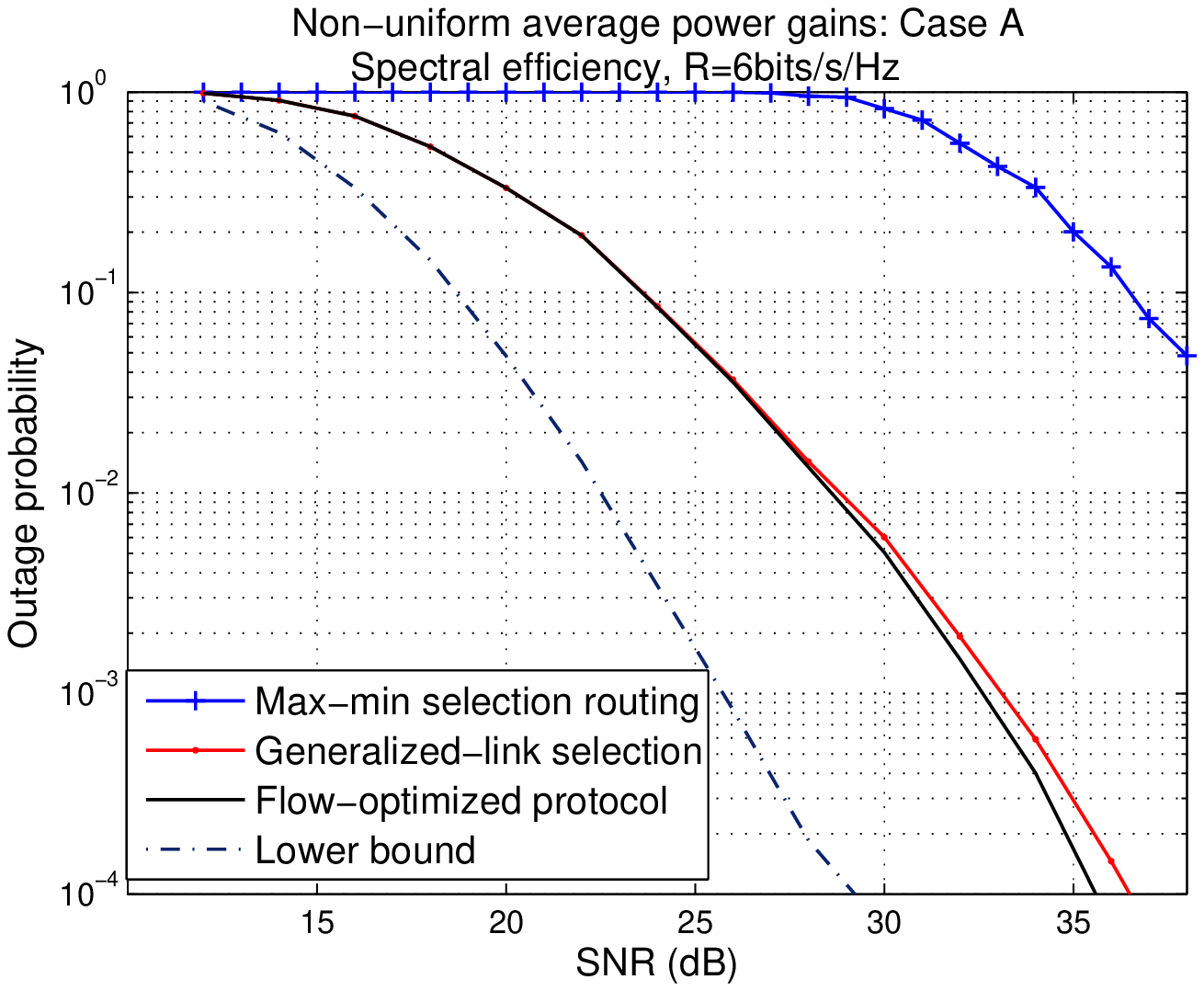}
\end{center}
\caption{Four-node relay network with non-uniform average power
gains. Case A:
$\mathbb{E}[Z_{\mathcal{SR}_1}]=2.0,~\mathbb{E}[Z_{\mathcal{SR}_2}]=2.0,~\mathbb{E}[Z_{\mathcal{SD}}]=1.0,
~\mathbb{E}[Z_{\mathcal{R}_1\mathcal{R}_2}]=1.0,~\mathbb{E}[Z_{\mathcal{R}_1\mathcal{D}}]=1.5,~\mathbb{E}[Z_{\mathcal{R}_2\mathcal{D}}]=1.0$.
In this case, both the source-relay links
are, on average, better than the direct link, and one
relay-destination link is, on average, better than the other,
resulting in relay $\mathcal{R}_1$ being more favorable than relay $\mathcal{R}_2$.}
\label{fig:case_6a_R6}
\end{figure}
\begin{figure}[t]
\begin{center}
\includegraphics[scale=1.0]{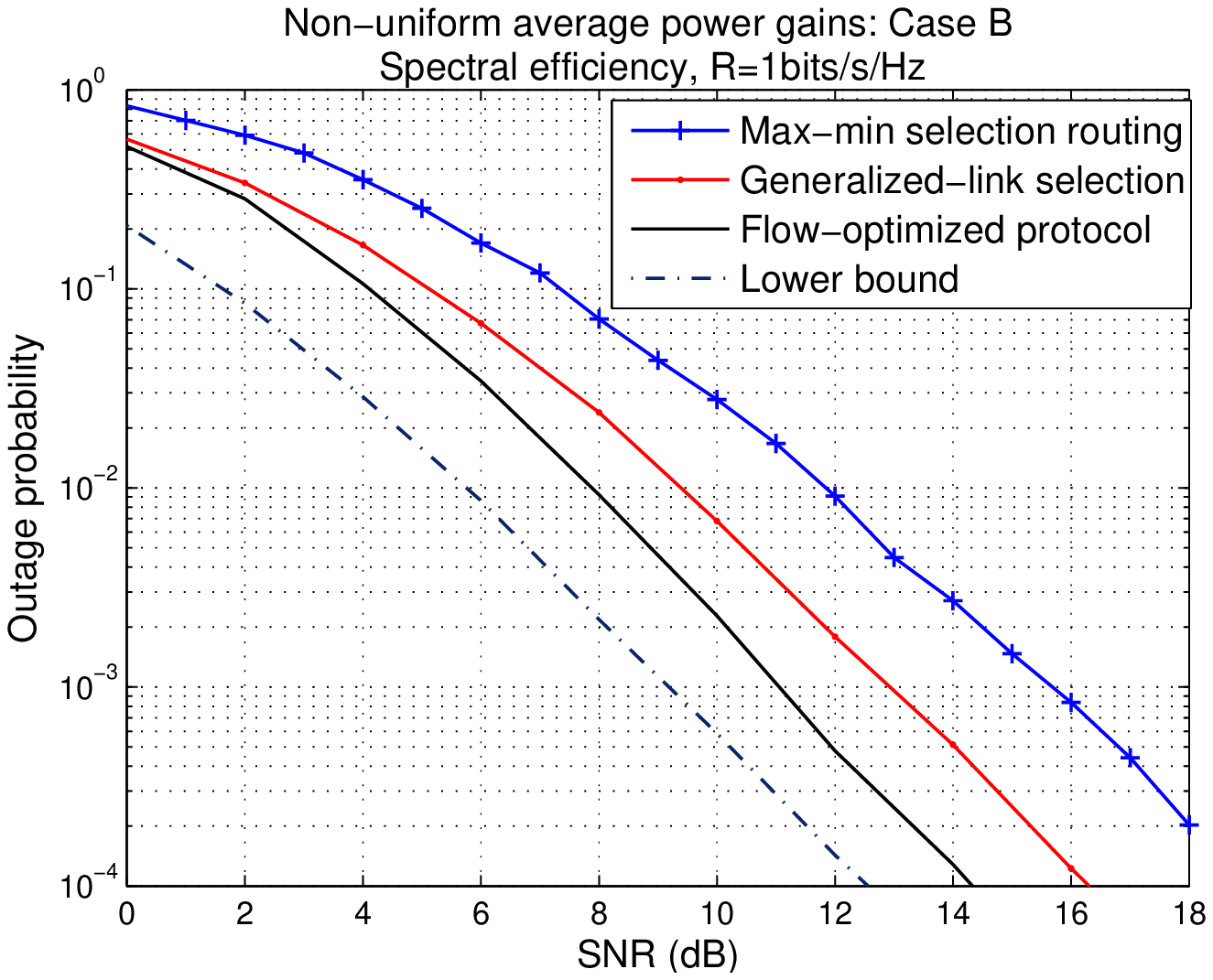}
\end{center}
\caption{Four-node relay network with non-uniform average power
gains. Case B:
$\mathbb{E}[Z_{\mathcal{SR}_1}]=1.5,~\mathbb{E}[Z_{\mathcal{SR}_2}]=0.75,~\mathbb{E}[Z_{\mathcal{SD}}]=1.0,
~\mathbb{E}[Z_{\mathcal{R}_1\mathcal{R}_2}]=3.5,~\mathbb{E}[Z_{\mathcal{R}_1\mathcal{D}}]=0.2,~\mathbb{E}[Z_{\mathcal{R}_2\mathcal{D}}]=3.0$.
In this case, one source-relay link is, on
average, better than the direct link, which, in turn, is better than
the other source-relay link. The reverse is true for the
relay-destination links, and the inter-relay channel is, on average,
very good. This situation promotes inter-relay interactions for the
FO protocol, and thereby highlights the sub-optimality of the GLS protocol.}
\label{fig:case_1b_R1}
\end{figure}
\begin{figure}[t]
\begin{center}
\includegraphics[scale=1.0]{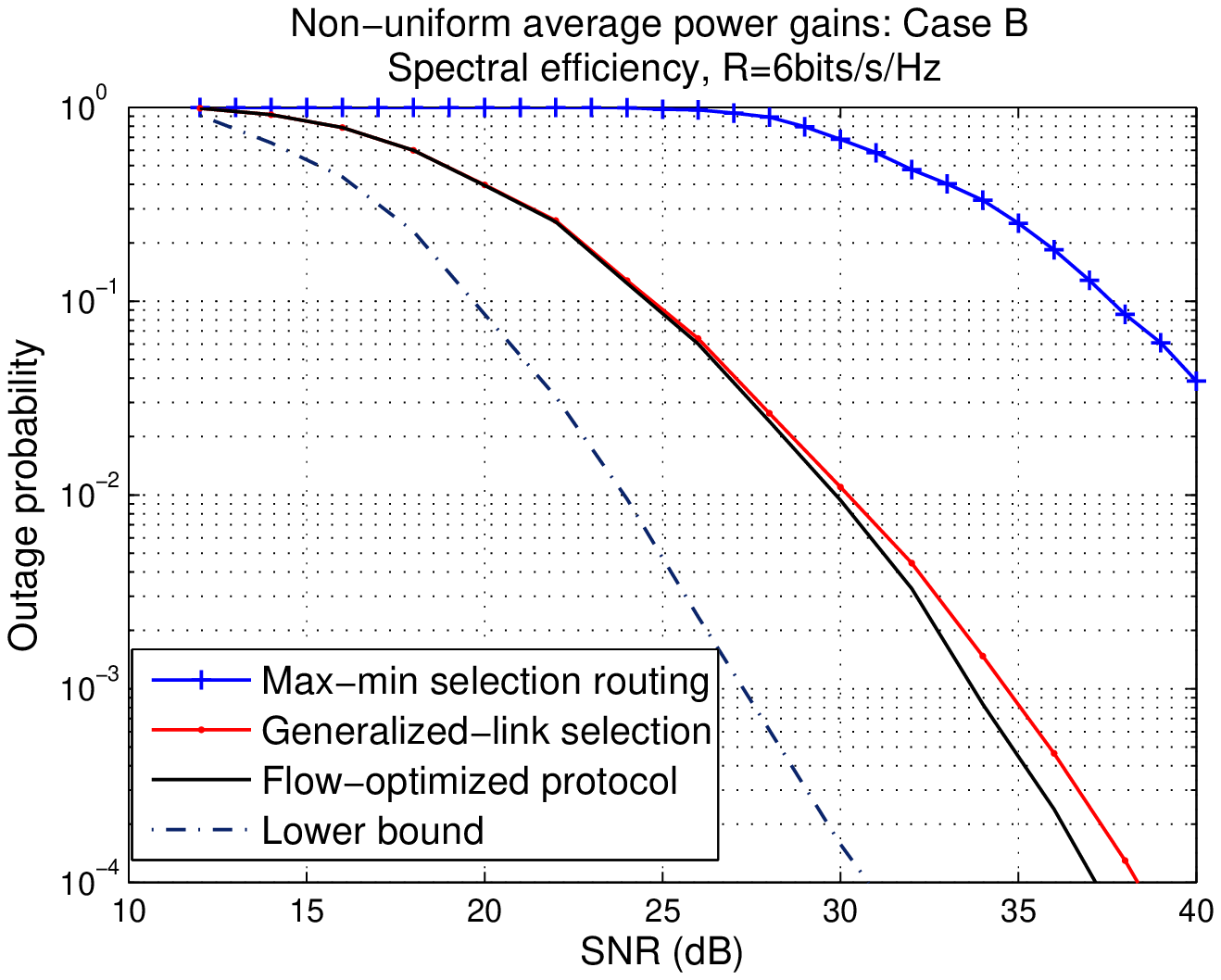}
\end{center}
\caption{Four-node relay network with non-uniform average power
gains. Case B:
$\mathbb{E}[Z_{\mathcal{SR}_1}]=1.5,~\mathbb{E}[Z_{\mathcal{SR}_2}]=0.75,~\mathbb{E}[Z_{\mathcal{SD}}]=1.0,
~\mathbb{E}[Z_{\mathcal{R}_1\mathcal{R}_2}]=3.5,~\mathbb{E}[Z_{\mathcal{R}_1\mathcal{D}}]=0.2,~\mathbb{E}[Z_{\mathcal{R}_2\mathcal{D}}]=3.0$.
In this case, one source-relay link is, on
average, better than the direct link, which, in turn, is better than
the other source-relay link. The reverse is true for the
relay-destination links, and the inter-relay channel is, on average,
very good. This situation promotes inter-relay interactions for the
FO protocol, and thereby highlights the sub-optimality of the GLS protocol.}
\label{fig:case_6b_R6}
\end{figure}

The performances of the different protocols for the four-node
relay network with non-uniform average power gains are presented in
Figs.~\ref{fig:case_1a_R1} and~\ref{fig:case_6a_R6}, and
Figs.~\ref{fig:case_1b_R1} and~\ref{fig:case_6b_R6} for cases A and
B respectively, with the average power gains as stated in the
figures. In case A, the
source-relay links are, on average, better than the direct link, and
one relay node is, on average, a better candidate to forward the
information to $\mathcal{D}$. On the other hand, in case B,
no one relay has very good source-relay and
relay-destination links, whereas the inter-relay link is, on
average, very good. This promotes increased inter-relay flows when
using the FO protocol, and thereby highlights the limitations of the
GLS protocol. The differences between the outage performances of the
FO and GLS protocols, at an outage probability of $10^{-4}$, are
$1.2$dB or $1.0$dB, and $2.0$dB or $1.3$dB (when $R=1$bits/s/Hz or
$R=6$bits/s/Hz), for cases A and B respectively. Thus, the gap between
the FO and GLS protocols decreases as the required
data rate increases. When the required rate is
high, the coding gain offered by a protocol heavily relies on the
efficient use of the direct link, and since the usage of the direct
link is similar for both the FO and the GLS protocols, the
performance gap narrows as the required data rate increases. On the
other hand, at the same outage probability, the difference between
the outage performance of the FO protocol and the lower bound
increases from $1.5$dB to $7$dB, and from $1.9$dB to $6.0$dB as the
required rate increases from $1$bit/s/Hz to $6$bits/s/Hz, for cases
A and B respectively. Overall, these results demonstrate trends
similar to the uniform average power gain case, and confirm the
generality of the proposed protocols.

\section{Conclusions}\label{sec:conclusions}

We proposed a cooperative transmission design for a general
multi-node half-duplex wireless relay network. 
It is based on optimizing information flows, using the basic
components of BC and MA, to maximize the transmission rate from the
source to the destination, subject to maximum power constraints at
individual nodes. We also proposed the simpler GLS protocol,
that combines relay selection, and flow
optimization for a three-node relay network.
These protocols were shown to achieve the optimal
DMT for a general relay network. Simulation results for the four- and five-node
relay networks for uniform and non-uniform average power gains demonstrate
that the performance of the much simpler GLS protocol is close to
 that of the FO protocol.
This suggests that the GLS protocol can be used in systems with
low-complexity requirements. We also note that the proposed FO and
GLS protocols can be used in wireless networks with topologies more
complicated that the wireless relay network considered here. For example,
application of similar ideas to a parallel relay network in which
there is no direct connection between the source and the destination
is considered in \cite{Tam07}.

\appendix

\subsection{Solution to optimization problem~\eqref{eqn:3node_opt}}
\label{app:opt_soln}

We consider two cases with regard to the link gains: (a)
$Z_{\mathcal{SD}}\geq Z_{\mathcal{SR}}$, and (b) $Z_{\mathcal{SD}}<
Z_{\mathcal{SR}}$. For both cases, we solve the optimization problem
in two stages: first, we fix $t_1,~t_2~\geq0$ such that $t_1+t_2=1$
and find the optimal flows $x_1,~x_2,~x_3$ in terms of $t_1,~t_2$,
and then, find the optimal values for $t_1,~t_2$ to maximize the
objective function.
\paragraph{$Z_{\mathcal{SD}}\geq Z_{\mathcal{SR}}$}
To obtain an analytical solution to the
optimization problem and better insight into the nature of the
solution to the flow optimization problem, we modify the
representation of the BC slot power constraint from that
in~\eqref{eqn:3node_opt} to the one that is more conventionally used to
describe the capacity region of the Gaussian BC channel, as
presented
in~\eqref{eqn:3node_opt_a}. Using the flow constraint in
(\ref{eqn:3node_opt}), we first solve~\eqref{eqn:3node_opt_a} for
fixed $t_1,~t_2$.
\begin{equation}
\displaystyle\max(x_1+x_2+x_3)~~\textnormal{over}~x_1,~x_2,~x_3,~\alpha,~\textnormal{subject to} \label{eqn:3node_opt_a}
\end{equation}
\vspace{-10mm}
\begin{eqnarray}
&&x_1,~x_2,~x_3\geq0,~~ 0\leq\alpha\leq1,\nonumber\\
&&x_1\leq t_1C(Z_{\mathcal{SD}}\alpha S),~~x_2\leq
t_1C\left(\frac{Z_{\mathcal{SR}}\bar{\alpha}
S}{1+Z_{\mathcal{SR}}\alpha S}\right),\nonumber\\
&&x_2\leq t_2C(Z_{\mathcal{RD}}S), ~~x_3\leq
\min\left\{t_2C(Z_{\mathcal{SD}}S),
~t_2C(Z_{\mathcal{SD}}S+Z_{\mathcal{RD}}S)- x_2\right\}\nonumber.
\end{eqnarray}
Here, $\alpha\in[0,1]$ is the fraction of total power spent at the
source to transmit $x_1$ directly to the destination during the BC
slot, and $\bar{\alpha}=1-\alpha$. Although, this modification of
the BC slot power constraint apparently makes the optimization
problem non-convex owing to the non-convexity in $\alpha$, as we
shall see in the sequel, this issue can be handled easily by
utilizing the monotonicity of the logarithm function.

Denote the optimal solution by $(x_1^*, x_2^*, x_3^*, \alpha^*)$ and
the corresponding maximum rate by $X(t_1,t_2)$. It is clear that
$x_1^* = t_1C(Z_{\mathcal{SD}}\alpha^* S)$. Suppose that
$x_2^*<t_1C\left(\frac{Z_{\mathcal{SR}}\bar{\alpha}^*
    S}{1+Z_{\mathcal{SR}}\alpha^* S}\right)$. Since
$t_1C\left(\frac{Z_{\mathcal{SR}}\bar{\alpha}
    S}{1+Z_{\mathcal{SR}}\alpha S}\right)$ is a decreasing function of
$\alpha$, we can increase $\alpha$ from $\alpha^*$ to $\alpha^0$ such
that $x_1^0=t_1C(Z_{\mathcal{SD}}\alpha^0 S)>x_1^*$ and
$x_2^*=t_1C\left(\frac{Z_{\mathcal{SR}}\bar{\alpha}^0
    S}{1+Z_{\mathcal{SR}}\alpha^0 S}\right)$. Thus the objective
function becomes $(x_1^0+x_2^*+x_3^*)>(x_1^*+x_2^*+x_3^*)$ at
$\alpha^0$. This contradicts the optimality of $(x_1^*, x_2^*,
x_3^*, \alpha^*)$. As a consequence, we have
$x_2^*=t_1C\left(\frac{Z_{\mathcal{SR}}\bar{\alpha}^*
S}{1+Z_{\mathcal{SR}}\alpha^* S}\right)\leq
t_2C(Z_{\mathcal{RD}}S)$. This implies that $\alpha^*\geq
\max\Bigg\{\underbrace{\frac{1}{Z_{\mathcal{SR}}S}\left[\frac{1+Z_{\mathcal{SR}}S}{(1+Z_{\mathcal{RD}}S)^{t_2/t_1}}-1\right]}_{\alpha_0^a},
0\Bigg\},~\textnormal{and}~\alpha_a^0\leq1.$ In essence, this means
that the optimal $x_1$ and $x_2$ should lie on the boundary of the
degraded BC capacity region. With this, it is obvious then that $x_3
= \min\left\{t_2C(Z_{\mathcal{SD}}S),
~t_2C(Z_{\mathcal{SD}}S+Z_{\mathcal{RD}}S)-
t_1C\left(\frac{Z_{\mathcal{SR}}\bar{\alpha}S}{1+Z_{\mathcal{SR}}\alpha
S}\right)\right\}$. Therefore the optimization problem of (\ref{eqn:3node_opt_a}) can be
re-written as:
\vspace{-5mm}
\begin{eqnarray}
\displaystyle\max&&(x_1+x_2+x_3)
\label{eqn:3node_opt_a1}\\
\textnormal{subject to}~~ &&\max\{0,\alpha_a^0\}\leq\alpha\leq1,~~
x_1 = t_1C(Z_{\mathcal{SD}}\alpha S), ~~ x_2 =
t_1C\left(\frac{Z_{\mathcal{SR}}\bar{\alpha}
S}{1+Z_{\mathcal{SR}}\alpha S}\right),\nonumber\\
&&x_3 = \min\left\{t_2C(Z_{\mathcal{SD}}S),
~t_2C(Z_{\mathcal{SD}}S+Z_{\mathcal{RD}}S)-
t_1C\left(\frac{Z_{\mathcal{SR}}\bar{\alpha}S}{1+Z_{\mathcal{SR}}\alpha
S}\right)\right\}\nonumber.
\end{eqnarray}
We observe that $x_3 = t_2C(Z_{\mathcal{SD}}S)$ above if and only if
\begin{equation}
\alpha\geq
\frac{1}{Z_{\mathcal{SR}}S}\left[\frac{1+Z_{\mathcal{SR}}S}{\left(1+\frac{Z_{\mathcal{RD}}S}{1+Z_{\mathcal{SD}}S}\right)^{t_2/t_1}}-1\right]\triangleq\alpha_a^1.\label{eqn:3node_alpha_a1}
\end{equation}
Comparing this to the expression for $\alpha_a^0$ gives
$\alpha_a^0\leq \alpha_a^1\leq1$.

\noindent Next we consider two possible sub-cases:
\begin{enumerate}
\item[i.]
  $t_2C\left(\frac{Z_{\mathcal{RD}}S}{1+Z_{\mathcal{SD}}S}\right) \leq
  t_1C(Z_{\mathcal{SR}}S)$: \\
  In this case, we have $\bar{\alpha}_a^1 \geq 0$ and $t_2 \leq
  \frac{C(Z_{\mathcal{SR}}S)}{C\left(\frac{Z_{\mathcal{RD}}S}%
      {1+Z_{\mathcal{SD}}S}\right)+C(Z_{\mathcal{SR}}S)}~
  \triangleq t_{2\textnormal{max}}$. The maximum rate can be expressed
  as $X(t_1,t_2)=\max\{X_1(t_1,t_2),~X_2(t_1,t_2)\}$ where
\begin{eqnarray}
X_1(t_1,t_2) &=& \max_{\max\{0,\alpha_a^0\} \leq
\alpha
\leq \alpha_a^1} ~t_1C(Z_{\mathcal{SD}}\alpha S) + t_2C(Z_{\mathcal{SD}}S+Z_{\mathcal{RD}}S)\nonumber\\
&=& t_1C(Z_{\mathcal{SD}}\alpha_a^1S) + t_2C(Z_{\mathcal{SD}}S+Z_{\mathcal{RD}}S)\nonumber\\
&\leq&
t_1\log\left(\frac{(1+Z_{\mathcal{SD}}\alpha_a^1S)(1+Z_{\mathcal{SR}}S)}{1+Z_{\mathcal{SR}}\alpha_a^1S}\right)+t_2C(Z_{\mathcal{SD}}S)\leq C(Z_{\mathcal{SD}}S)\label{eqn:3node_aX1}\\
X_2(t_1,t_2) &=&
\max_{\alpha_a^1 \leq \alpha \leq 1} ~t_1C(Z_{\mathcal{SD}}\alpha
S)+t_1C\left(\frac{Z_{\mathcal{SR}}\bar{\alpha}S}{1+Z_{\mathcal{SR}}\alpha
S}\right)+t_2C(Z_{\mathcal{SD}}S)\nonumber\\
&=& C(Z_{\mathcal{SD}}S)\label{eqn:3node_aX2}
\end{eqnarray}
where the first inequality in (\ref{eqn:3node_aX1}) holds since
(\ref{eqn:3node_alpha_a1}) is not satisfied, and the second
inequality in (\ref{eqn:3node_aX1}) holds since
$\alpha_a^1\leq1$ and that the first term in the previous step is
monotonically increasing in $\alpha_a^1$ when $Z_{\mathcal{SD}}\geq
Z_{\mathcal{SR}}$. This way, the last observation helps avoid the
non-convexity issue mentioned before.
Similar arguments hold for $X_2(t_1,t_2)$ in~\eqref{eqn:3node_aX2}.

\item[ii.] $t_2C\left(\frac{Z_{\mathcal{RD}}S}{1+Z_{\mathcal{SD}}S}\right) >
t_1C(Z_{\mathcal{SR}}S)$:\\
\noindent In this case, we have $\alpha_a^1 < 0$ and $t_2
> t_{2\textnormal{max}}$. 
Thus the maximum rate is given by
\begin{equation}
X(t_1,t_2) =
\max_{0\leq\alpha\leq1}~t_1C(Z_{\mathcal{SD}}\alpha
S)+t_1C\left(\frac{Z_{\mathcal{SR}}\bar{\alpha}S}{1+Z_{\mathcal{SR}}\alpha
S}\right)+t_2C(Z_{\mathcal{SD}}S)
= C(Z_{\mathcal{SD}}S).\label{eqn:3node_aX2ii}
\end{equation}
\end{enumerate}

\noindent Hence,~\eqref{eqn:3node_aX1}--\eqref{eqn:3node_aX2ii}
imply that the solution to (\ref{eqn:3node_opt_a}), for any
$t_1,~t_2$ pair, occurs at $\alpha=1$, and the solution to the
original problem of (\ref{eqn:3node_opt}), when
$Z_{\mathcal{SD}}\geq Z_{\mathcal{SR}}$, is given by $\max_{\{0\leq
t_1,t_2~:~t_1+t_2=1\}}~X(t_1,t_2) = C(Z_{\mathcal{SD}}S)$ with any
$t_1, t_2$ pair such that $t_1,t_2\geq0$ and $t_1+t_2=1$.

\paragraph{$Z_{\mathcal{SD}} < Z_{\mathcal{SR}}$}
In this case, the source-relay link is better than the
source-destination link. Again, we first fix time slot lengths $t_1$
and $t_2$ and solve for the optimal values of $x_1,~x_2,~x_3$, and
$\alpha$, and then maximize the objective function of
(\ref{eqn:3node_opt}) over all feasible time slot lengths.  Following
similar arguments as in Case a), the optimization problem of
(\ref{eqn:3node_opt}) can be re-written as:
\begin{eqnarray}
\max&&(x_1+x_2+x_3) \label{eqn:3node_opt_b1}\\
\textnormal{subject to}~~
&&0\leq\bar{\alpha}\leq\min\{1,\bar{\alpha}_b^0\},~~x_1 =
t_1C\left(\frac{Z_{\mathcal{SD}}\alpha
S}{1+Z_{\mathcal{SD}}\bar{\alpha}S}\right), ~~ x_2 =
t_1C(Z_{\mathcal{SR}}\bar{\alpha}S),\nonumber\\
&&x_3 = \min\left\{t_2C(Z_{\mathcal{SD}}S),
~t_2C(Z_{\mathcal{SD}}S+Z_{\mathcal{RD}}S)-
t_1C(Z_{\mathcal{SR}}\bar{\alpha}S)\right\}\nonumber
\end{eqnarray}
where
$\bar{\alpha}_b^0=\frac{1}{Z_{\mathcal{SR}}S}\left[(1+Z_{\mathcal{RD}}S)^{t_2/t_1}-1\right]$
is an upper bound on $\bar{\alpha}$ such that
$t_1C(Z_{\mathcal{SR}}\bar{\alpha}S)\leq t_2C(Z_{\mathcal{RD}}S)$.
As before, $x_3=t_2C(Z_{\mathcal{SD}}S)$ if and only if
$\bar{\alpha} \leq
\frac{1}{Z_{\mathcal{SR}}S}\left[\left(1+\frac{Z_{\mathcal{RD}}S}{1+Z_{\mathcal{SD}}S}\right)^{t_2/t_1}-1\right]\triangleq\bar{\alpha}_b^1$.
Also, $0\leq\bar{\alpha}_b^1\leq\bar{\alpha}_b^0$. Again, we
consider two possible sub-cases:
\begin{enumerate}
\item[i.] $t_2C\left(\frac{Z_{\mathcal{RD}}S}{1+Z_{\mathcal{SD}}S}\right) \leq
t_1C(Z_{\mathcal{SR}}S)$:\\
\noindent In this case we have $\bar{\alpha}_b^1 \leq 1$ and $t_2
\leq
\frac{C(Z_{\mathcal{SR}}S)}{C\left(\frac{Z_{\mathcal{RD}}S}{1+Z_{\mathcal{SD}}S}\right)+C(Z_{\mathcal{SR}}S)}~
\triangleq t_{2\textnormal{max}}$. The maximum rate $X(t_1,t_2)$ can
be expressed as $X(t_1,t_2) = \max\{X_1(t_1,t_2),~X_2(t_1,t_2)\}$
where
\begin{eqnarray}
X_1(t_1,t_2) &=&
\max_{0\leq\bar{\alpha}\leq\bar{\alpha}_b^1}~t_1C\left(\frac{Z_{\mathcal{SD}}\alpha
S}{1+Z_{\mathcal{SD}}\bar{\alpha}S}\right)+t_1C(Z_{\mathcal{SR}}\bar{\alpha}S)+t_2C(Z_{\mathcal{SD}}S),
\label{eqn:3node_bX1}\\
X_2(t_1,t_2) &=& \max_{\bar{\alpha}_b^1 \leq \bar{\alpha} \leq
\min\{1,\bar{\alpha}_b^0\}} ~ t_1C\left(\frac{Z_{\mathcal{SD}}\alpha
S}{1+Z_{\mathcal{SD}}\bar{\alpha}S}\right) +
t_2C(Z_{\mathcal{SD}}S+Z_{\mathcal{RD}}S)
,\label{eqn:3node_bX2}
\end{eqnarray}
and both the maxima in~\eqref{eqn:3node_bX1}
and~\eqref{eqn:3node_bX2} are attained at
$\bar{\alpha}=\bar{\alpha}_b^1$. Substituting the expression for
$\bar{\alpha}_b^1$ in (\ref{eqn:3node_bX1}) and
(\ref{eqn:3node_bX2}), we obtain $X_1(t_1,t_2) = X_2(t_1,t_2)$ and
\begin{equation}
\displaystyle X(t_1,t_2) = t_1\log\left(\frac{1+Z_{\mathcal{SD}}S}
{1+\frac{Z_{\mathcal{SD}}}{Z_{\mathcal{SR}}}\left[\left(1+\frac{Z_{\mathcal{RD}}S}{1+Z_{\mathcal{SD}}S}\right)^{t_2/t_1}-1\right]}\right)
+ t_2C(Z_{\mathcal{SD}}S+Z_{\mathcal{RD}}S).\label{eqn:3node_iX}
\end{equation}

\item[ii.] $t_2C\left(\frac{Z_{\mathcal{RD}}S}{1+Z_{\mathcal{SD}}S}\right) >
t_1C(Z_{\mathcal{SR}}S)$:\\
\noindent In this case we have $\bar{\alpha}_b^1 > 1$ and $t_2
> t_{2\textnormal{max}}$. 
Thus the maximum rate is given by
\begin{eqnarray}
X(t_1,t_2) &=&
\max_{0\leq\bar{\alpha}\leq1}~t_1C\left(\frac{Z_{\mathcal{SD}}\alpha
S}{1+Z_{\mathcal{SD}}\bar{\alpha}S}\right)+t_1C(Z_{\mathcal{SR}}\bar{\alpha}S)+t_2C(Z_{\mathcal{SD}}S)\nonumber\\
&=& t_1C(Z_{\mathcal{SR}}S) +
t_2C(Z_{\mathcal{SD}}S)\label{eqn:3node_iiX}
\end{eqnarray}
\noindent where the maximum occurs at $\bar{\alpha}=1$, as the
function to be maximized is monotonically increasing in
$\bar{\alpha}$ when $Z_{\mathcal{SR}}>Z_{\mathcal{SD}}$. 
\end{enumerate}

\noindent Finally, we optimize the above solution to
(\ref{eqn:3node_opt_b1}) over all possible time slot lengths to
obtain the solution to the original problem in (\ref{eqn:3node_opt})
when $Z_{\mathcal{SD}}<Z_{\mathcal{SR}}$. Corresponding to Case i.
above, when $t_2\leq t_{2\textnormal{max}}$, we note that
\begin{eqnarray}
\max_{\{0\leq t_1,t_2:t_1+t_2=1\}}X(t_1,t_2) &\geq&
X(1-t_2,t_2)|_{t_2=t_{2\textrm{max}}}\nonumber\\
&=& \frac{C(Z_{\mathcal{SR}}S)
C(Z_{\mathcal{SD}}S+Z_{\mathcal{RD}}S)}{C(Z_{\mathcal{SR}}S)+C(Z_{\mathcal{RD}}S+Z_{SD}S)-C(Z_{\mathcal{SD}}S)}.\label{eqn:case_i_X}
\end{eqnarray}
On the other hand, corresponding to Case ii., when $t_2>
t_{2\textnormal{max}}$, from (\ref{eqn:3node_iiX}), we have
\begin{eqnarray}
\displaystyle\max_{\{0\leq t_1,t_2:~t_1+t_2=1\}}X(t_1,t_2) &<&
(1-t_{2\textrm{max}})C(Z_{\mathcal{SR}}S)+t_{2\textrm{max}}C(Z_{\mathcal{SD}}S)\nonumber\\
&=& \frac{C(Z_{\mathcal{SR}}S)
C(Z_{\mathcal{SD}}S+Z_{\mathcal{RD}}S)}{C(Z_{\mathcal{SR}}S)+C(Z_{\mathcal{RD}}S+Z_{SD}S)-C(Z_{\mathcal{SD}}S)}\label{eqn:case_ii_X}
\end{eqnarray}
\noindent where the inequality in (\ref{eqn:case_ii_X}) is obtained
from (\ref{eqn:3node_iiX}) by using
$Z_{\mathcal{SR}}>Z_{\mathcal{SD}}~\text{and}~t_2 >
t_{2\textrm{max}}$.  Hence, from (\ref{eqn:case_i_X}) and
(\ref{eqn:case_ii_X}), we conclude that when
$Z_{\mathcal{SR}}>Z_{\mathcal{SD}}$, the maximum achievable rate is
given by $X(S) =\max_{0\leq t_2\leq t_{2\textnormal{max}}}
t_1\log\left(\frac{1+Z_{\mathcal{SD}}S}
{1+\frac{Z_{\mathcal{SD}}}{Z_{\mathcal{SR}}}\left[\left(1+\frac{Z_{\mathcal{RD}}S}{1+Z_{\mathcal{SD}}S}\right)^{t_2/t_1}-1\right]}\right)
+t_2C(Z_{\mathcal{SD}}S+Z_{\mathcal{RD}}S)$.

\subsection{Proof of Theorem \ref{thm:dm_sub_opt}}
\label{app:pf_dm_sub_opt}
In this appendix, we sketch the proof of the theorem.  Part 3) of
Theorem 4.2 of \cite{Wong06} can be generalized for the $N$-node
relay network to prove that, as the block length goes to infinity
(during any particular time interval), the average error probability
for the FO protocol is upper bounded by its corresponding outage
probability. Here, the outage probability denotes the probability that
the data rate $R$ cannot be supported by the system when the SNR is
$S$, i.e., $P_{\textnormal{out}}(r,S) = \Pr [ X(S) < r\log S ]$ where
$X(S)$ denotes the maximum rate possible for the given channel gain
realizations when the SNR is $S$. Thus, from the definition of
diversity order (\ref{eqn:div_def}), we have
\begin{equation}
\displaystyle d(r)\geq\lim_{S \rightarrow \infty} \frac{-\log
P_{\textnormal{out}} (r,S)}{\log S}.\label{eqn:divord_lb}
\end{equation}
Moreover, the above result from \cite{Wong06} can be directly used
to prove the same for the GLS protocol.  Using this fact, we derive
a lower bound to the DMT that can be
achieved by the GLS protocol. The sets $I$ and $K$, used in the
sequel, are the sets of indices as described in Section
\ref{subsec:sub_opt}. The outage probability for the GLS protocol is
given by
\begin{equation}
P_{\textnormal{out}}^{\textnormal{gr}}(r,S)=
\Pr\left(\max_{k\in I}X_k(S)<r\log S\right)\label{eqn:outage_gr_def}
\end{equation}
where $X_k(S)$ is the maximum rate achievable by the three-node
relay network formed by the source $\mathcal{S}$, the relay $\mathcal{R}_k$
and the destination $\mathcal{D}$. We have the following
possibilities:
\begin{itemize}
\item Case A: $|K|=0$, i.e. the cardinality of the set $K$ is zero.
This corresponds to the case when $Z_{\mathcal{SD}}\geq
Z_{\mathcal{SR}_k}$ for all $k\in I$.
\item Case B: $|K|=i$ with $i\in I$.
\end{itemize}
Note that for Case B there are $N-2\choose i$ possibilities for the
set $K$ with cardinality $i$. Since the link gains are assumed to be
i.i.d., and the outage probability depends on the distribution of the
maximum of $X_k(S)$ over all $k\in I$ (or effectively, over all $k\in
K$ when $|K|>0$), only the cardinality of $K$ is significant. Let the
$N-2\choose i$ possible constructions of the set $K$ be represented by
a ``generic'' set $K_0$ with cardinality $i$. Without loss of
generality, we describe $K_0$ as the set $K$ corresponding to the case
when the indices of the relay nodes are ordered according to their
source-relay link gains, i.e.  $Z_{\mathcal{SR}_1}\geq
Z_{\mathcal{SR}_2}\geq\cdots\geq Z_{\mathcal{SR}_{N-2}}$. Thus, Case B
now implies a solitary choice for set $K$, viz.
$K_0=\{1,2,\cdots,i\}$. Therefore, from (\ref{eqn:outage_gr_def}), we
have
\begin{eqnarray}
P_{\textnormal{out}}^{\textnormal{gr}}(r,S)&=&
\Pr\left(C(Z_{\mathcal{SD}}S)<r\log S \middle|
~|K_0|=0\right)\Pr(|K_0|=0)\nonumber\\
&&+\sum_{i=1}^{N-2}\Pr\left(\max_{k\in K_0}X_k(S)<r\log S\middle|
~|K_0|=i\right)\Pr(|K_0|=i).\label{eqn:outage_gr}
\end{eqnarray}
We observe that using the right-most expression of
(\ref{eqn:case_i_X}) instead of $X_k(S)$, for each $k\in K_0$, in
(\ref{eqn:outage_gr}) gives an upper bound on
$P_{\textnormal{out}}^{\textnormal{gr}}(r,S)$. This is utilized in
obtaining a lower bound on the diversity order of the GLS protocol.
Let $\{S_n\}_{n=1}^{\infty}$ be an increasing unbounded sequence of
SNRs with $S_1>1$. Define the sequence of random variables
$\{M_n^k\}_{n=1}^{\infty}$, $\{B_n^k\}_{n=1}^{\infty}$ and
$\{A_n^k\}_{n=1}^{\infty}$ with $M_n^k =
\frac{C(Z_{\mathcal{SR}_k}S_n)-C(Z_{\mathcal{SD}}S_n)}%
{C(Z_{\mathcal{SD}}S_n+Z_{\mathcal{R}_k\mathcal{D}}S_n)}$, $B_n^k =
\frac{C(Z_{\mathcal{SR}_k}S_n)}{\log S_n}$, and $A_n^k =
\frac{X_k(S_n)}{\log S_n} = \frac{B_n^k}{1+M_n^k}$, respectively.
Note that for all $k\in K_0$, $M_n^k \rightarrow 0$ a.s.  This implies
that $(B_n^k - A_n^k)\rightarrow 0$ a.s.  Define $A'_{n} = \max_{k\in
  K_0} A_n^k$ and $B'_{n} = \max_{k\in K_0} B_n^k~(=B_n^1)$. Then
using the above, it can be seen that $(B'_{n}-A'_{n}) \rightarrow 0$
a.s.  Further,
$\displaystyle\lim_{n\to\infty}\Pr\left(B'_{n}<r\middle|~
  |K_0|=i\right)$ exists, and therefore the above implies that
$\displaystyle\lim_{n\to\infty}\Pr\left(A'_{n}<r\middle|~
  |K_0|=i\right)=\lim_{n\to\infty}\Pr\left(B'_{n}<r\middle|~
  |K_0|=i\right)$. Using this in (\ref{eqn:divord_lb}) and
(\ref{eqn:outage_gr}), the diversity order for the GLS protocol,
$d^{\textnormal{gr}}(r)$, satisfies
\begin{eqnarray}
\displaystyle
d^{\textnormal{gr}}(r) &\geq& \lim_{S \rightarrow
\infty} \frac{-\log P_{\textnormal{out}}^{\textnormal{gr}}(r,S)}{\log S}\nonumber\\
&\geq&\lim_{n\to\infty} \frac{-1}{\log
S_n}\log\left[\Pr\left(Z_{\mathcal{SD}}<\frac{S_n^r-1}{S_n} \middle|
~|K_0|=0\right)\Pr(|K_0|=0)\right.\nonumber\\
&&\left. + \sum_{i=1}^{N-2}\Pr\left(
Z_{\mathcal{SR}_1}<\frac{S_n^r-1}{S_n} \middle|~|K_0|=i\right)\Pr(|K_0|=i)\right]\label{eqn:div_ord1}\\
&=&\lim_{n\to\infty} \frac{-1}{\log
S_n}\log\left[\Pr\left(\max\{Z_{\mathcal{SD}},Z_{\mathcal{SR}_1},\cdots,Z_{\mathcal{SR}_{N-2}}\}<\frac{S_n^r-1}{S_n}
\middle|
~|K_0|=0\right)\Pr(|K_0|=0)\right.\nonumber\\
&&\left. + \sum_{i=1}^{N-2}\Pr\left(\max\{Z_{\mathcal{SD}},Z_{\mathcal{SR}_1},\cdots,Z_{\mathcal{SR}_{N-2}}\}<\frac{S_n^r-1}{S_n}
\middle|~|K_0|=i\right)\Pr(|K_0|=i)\right]\label{eqn:div_ord2}\\
&=& \lim_{S\to\infty}
\frac{-\log\left(\Pr\left(\max\{Z_{\mathcal{SD}},Z_{\mathcal{S}\mathcal{R}_1},\cdots,Z_{\mathcal{S}\mathcal{R}_{N-2}}\}
< \frac{S^r-1}{S} \right)\right)}{\log S}\nonumber\\
&=& \lim_{S\to\infty} \frac{-\log\left(\left[1 -
\exp\left(-(S^r-1)/S\right)\right]^{N-1}\right)}{\log
S}=(N-1)(1-r)\label{eqn:d_genrout}
\end{eqnarray}
where (\ref{eqn:div_ord2}) is obtained from (\ref{eqn:div_ord1}) by
noting that
$\max\{Z_{\mathcal{SD}},Z_{\mathcal{SR}_1},\cdots,Z_{\mathcal{SR}_{N-2}}\}=Z_{\mathcal{SD}}$
when $|K_0|=0$, and
$\max\{Z_{\mathcal{SD}},Z_{\mathcal{SR}_1},\cdots,Z_{\mathcal{SR}_{N-2}}\}=\max_{k\in
  K_0}Z_{\mathcal{SR}_k}=Z_{\mathcal{SR}_1}$ when $|K_0|>0$, the first
equality in (\ref{eqn:d_genrout}) is due to the link gains being
i.i.d., and the second equality in (\ref{eqn:d_genrout}) is obtained
by using L'Hospital's rule.

Next given an $N$-node relay network, consider the multiple access cut that
separates the destination from all the other nodes.  Clearly, the
total flow across this cut gives an upper bound on the maximum rate
achievable in the $N$-node relay network. Consequently, a lower bound on the
outage probability $P_{\textnormal{out}}^l(r,S)$ can be obtained using
the maximum sum-rate across this cut:
\begin{eqnarray}
P_{\textrm{out}}^l(r,S) &\geq& \Pr \left[
C((Z_{\mathcal{SD}}+Z_{\mathcal{R}_1\mathcal{D}}+\cdots+Z_{\mathcal{R}_{N-2}\mathcal{D}})S) < r\log S\right]\nonumber\\
&=& \Pr
\left[(Z_{\mathcal{SD}}+Z_{\mathcal{R}_1\mathcal{D}}+\cdots+Z_{\mathcal{R}_{N-2}\mathcal{D}})
< x\right]=
\frac{\gamma(N-1,(S^r-1)/S)}{\Gamma(N-1)},\label{eqn:outprob_lower}
\end{eqnarray}
\noindent where $\gamma(a,x)=\int_0^{x}t^{a-1}e^{-t}dt$ is the lower
incomplete gamma function and
$\Gamma(a)=\int_0^{\infty}t^{a-1}e^{-t}dt$ is the complete gamma
function. The result in part 1) of Theorem 4.2 of \cite{Wong06} can
be extended to show that the diversity order of any transmission
scheme over the wireless relay network must satisfy
\begin{eqnarray}
\displaystyle d(r) &\leq& \lim_{S\to\infty}
\frac{-\log\left(P_{\textrm{out}}^l(r,S)\right)}{\log
S}
\leq  \lim_{S\to\infty}
\frac{-\log\left(\frac{\gamma\left(N-1,(S^r-1)/S\right)}{(N-2)!}\right)}{\log
S}\nonumber\\
&=& -\lim_{S\to\infty}
\frac{e^{-x}x^{N-2}[S^{-1}-(1-r)S^{r-1}]}{\int_0^xt^{N-2}e^{-t}dt}
=(N-1)(1-r).\label{eqn:u_bound}
\end{eqnarray}

\noindent Finally, from (\ref{eqn:d_genrout}) and
(\ref{eqn:u_bound}), we see that the GLS protocol, and hence the FO
protocol, achieve the optimal DMT of
$(N-1)(1-r)$ for all $0<r<1$.

\clearpage


\end{document}